\documentclass[journal,twoside,web]{ieeecolor}
\usepackage{generic}
\usepackage{cite}
\usepackage{textcomp}

\usepackage{amssymb,latexsym,amsfonts,amsmath,bbm}
\usepackage[cal=cm,bb=pazo]{mathalfa}
\usepackage{mathrsfs}
\usepackage{graphicx}
\usepackage{paralist}
\usepackage{dsfont}
\usepackage{eso-pic}
\usepackage{epsfig}
\usepackage{mathtools}
\usepackage{epstopdf}
\usepackage{lipsum}
\usepackage{cite}
\usepackage{caption}
\usepackage{subcaption}

\usepackage{booktabs}
\usepackage{multirow}

\usepackage[nocomma]{optidef}

\usepackage{tikz}
\usetikzlibrary{calc,shapes,arrows}

\usepackage{algorithm}
\usepackage{algorithmic}

\newcounter{theorem}
\newcounter{definition}
\newcounter{lemma}
\newcounter{claim}
\newcounter{problem}
\newcounter{proposition}
\newcounter{corollary}
\newcounter{construction}
\newcounter{example}
\newcounter{xca}
\newcounter{comments}
\newcounter{remark}
\newcounter{assumption}

\newtheorem{theorem}[theorem]{Theorem}
\newtheorem{lemma}[lemma]{Lemma}

\newtheorem{problem}[problem]{Problem}

\newtheorem{corollary}[corollary]{Corollary}

\newtheorem{definition}[definition]{Definition}

\newtheorem{remark}[remark]{Remark}
\newtheorem{assumption}[assumption]{Assumption}

\numberwithin{equation}{section}

\DeclareFontFamily{U}{stix2bb}{}
\DeclareFontShape{U}{stix2bb}{m}{n} {<-> stix2-mathbb}{}

\NewDocumentCommand{\stixbbdigit}{m}{%
	\text{\usefont{U}{stix2bb}{m}{n}#1}%
}
\newcommand{\bbzero}{\stixbbdigit{0}}

\usepackage[many]{tcolorbox}
\usetikzlibrary{calc}
\tcbuselibrary{skins}

\newtcolorbox{resp}[1][]{%
	enhanced jigsaw,%
	colback=gray!5!white,%
	colframe=gray!80!black,%
	size=small,%
	boxrule=1pt,%
	halign title=flush center,%
	coltitle=black,%
	breakable,%
	drop shadow=black!50!white,%
	attach boxed title to top left={xshift=1cm,yshift=-\tcboxedtitleheight/2,yshifttext=-\tcboxedtitleheight/2},%
	minipage boxed title=3cm,%
	boxed title style={%
		colback=white,%
		size=fbox,%
		boxrule=1pt,%
		boxsep=2pt,%
		underlay={%
			\coordinate (dotA) at ($(interior.west) + (-0.5pt,0)$);
			\coordinate (dotB) at ($(interior.east) + (0.5pt,0)$);
			\begin{scope}[gray!80!black]
				\fill (dotA) circle (2pt);
				\fill (dotB) circle (2pt);
			\end{scope}
		}%
	},%
	#1%
}

\usepackage{xspace}

\definecolor{blush}{rgb}{0.87, 0.36, 0.51}
\newcommand{\BS}[1]{\textcolor{black}{#1}}

\newcommand{\R}{{\mathbb{R}}}
\newcommand{\Rp}{{\mathbb{R}^+}}
\newcommand{\Rpz}{{\mathbb{R}^{+}_0}}
\newcommand{\N}{{\mathbb{N}}}
\newcommand{\Np}{{\mathbb{N}^+}}
\newcommand{\I}{{\mathbb{I}}}
\newcommand{\CT}{{\mathbf{\Sigma}}}
\newcommand{\CTROM}{{\mathbf{\hat{\Sigma}}}}
\newcommand{\CTSF}{{\mathbfcal{S}}}
\newcommand{\UC}{{\pmb{\mathbb{U}}}}
\newcommand{\XC}{{\pmb{\mathbb{X}}}}
\newcommand{\EC}{{\pmb{\mathbb{E}}}}
\newcommand{\MC}{{\pmb{\mathscr{M}}}}
\newcommand{\XdC}{{\pmb{\tilde{\mathbb{X}}^{\! +}}}}
\newcommand{\XdCm}{{\pmb{\mathbb{X}^{\! +}}}}
\newcommand{\DC}{{\pmb{\mathbb{D}}}}

\newcommand{\One}{{\mathbb{1}}}

\def\BibTeX{{\rm B\kern-.05em{\sc i\kern-.025em b}\kern-.08em
		T\kern-.1667em\lower.7ex\hbox{E}\kern-.125emX}}

\usepackage{etoolbox}

\makeatletter
\AtBeginDocument{%
	\pagestyle{headings}
	
	\patchcmd{\@oddhead}{\\[-19pt]}{\\[-8pt]}{}{}%
	\patchcmd{\@evenhead}{\\[-19pt]}{\\[-8pt]}{}{}%
}
\makeatother

\definecolor{blue(ryb)}{rgb}{0.01, 0.28, 1.0}
\definecolor{fashionfuchsia}{rgb}{0.96, 0.0, 0.63}
\definecolor{byzantine}{rgb}{0.74, 0.2, 0.64}
\definecolor{electricviolet}{rgb}{0.56, 0.0, 1.0}
\makeatletter
\let\NAT@parse\undefined
\makeatother
\usepackage[colorlinks=true, citecolor=electricviolet, linkcolor=electricviolet, final]{hyperref}
\renewcommand{\theequation}{\arabic{equation}}
\counterwithout*{equation}{section}

\providecommand\phantomsection{}

\makeatletter
\def\@opargbegintheorem#1#2#3{\textit{#1\ #2} \textit{(#3):}}
\makeatother

\DeclareRobustCommand{\legendsquare}[1]{%
	\textcolor{#1}{\rule{1.5ex}{1.5ex}}%
}

\definecolor{START}{rgb}{0.20 0.85 0.65}
\definecolor{TARGET}{rgb}{0.20 0.55 0.95}
\definecolor{OBSTACLES}{rgb}{1.0 0.45 0.45}

\definecolor{START1}{rgb}{0, 1, 0}
\definecolor{TARGET1}{rgb}{0.3010 0.7450 0.9330}
\definecolor{OBSTACLES1}{rgb}{1, 0, 0}

\usepackage{microtype}
\allowdisplaybreaks

\begin{document}
	
	\title{Data-Driven Model Order Reduction of Nonlinear Systems {with Noisy Data}}
	\author{Behrad Samari, \IEEEmembership{Student Member,~IEEE}, Henrik Sandberg, \IEEEmembership{Fellow,~IEEE,}
		\\ Karl H. Johansson, \IEEEmembership{Fellow,~IEEE}, and Abolfazl Lavaei, \IEEEmembership{Senior Member,~IEEE}
		\thanks{B. Samari and A. Lavaei are with the School of Computing, Newcastle University, NE4 5TG Newcastle Upon Tyne, United Kingdom (e-mails: {\tt\small{\{b.samari2,abolfazl.lavaei\}@newcastle.ac.uk}}).}
		\thanks{H. Sandberg and K. H. Johansson are with the Division of Decision and Control Systems, KTH Royal Institute of Technology, SE-100 44 Stockholm, Sweden. They are also affiliated with Digital Futures (e-mails: {\tt\small{\{hsan,kallej\}@kth.se}}).}
	}
	
	\maketitle
	\begin{abstract}
		Model order reduction techniques simplify high-dimensional dynamical systems by deriving lower-dimensional models that retain essential system characteristics. These techniques are crucial for the controller design of complex systems while significantly reducing computational costs. Nevertheless, constructing effective reduced-order models (ROMs) poses considerable challenges, particularly for nonlinear dynamical systems. These challenges are further exacerbated when the actual system model is unavailable, a scenario frequently encountered in real-world applications.
		In this work, we propose a \emph{data-driven} framework for constructing ROMs of nonlinear dynamical systems with unknown mathematical models, {enabling controller synthesis directly from the resulting ROMs.} We establish similarity relations between the output trajectories of the original systems and those of their ROMs by employing the notion of simulation functions (SFs), thereby enabling a formal characterization of their closeness.
		\BS{To achieve this, we collect one set of \emph{noise-corrupted} input--state data from the system during a finite-time experiment, upon which we propose conditions to construct both ROMs and SFs simultaneously. These conditions are formulated as \emph{data-dependent} semidefinite programs.} We demonstrate that the data-driven ROMs obtained can be employed to synthesize controllers for the original unknown systems, ensuring that they satisfy high-level logic specifications. This is accomplished by first designing controllers for the data-driven ROMs and then translating the results back to the original systems via interface functions, {designed directly from the proposed data-dependent conditions.} We evaluate the efficacy of our data-driven framework through {two} case studies, \BS{including a challenging benchmark from the model reduction literature: a circuit of chained inverter gates with 20 state variables.}
	\end{abstract}
	
	\begin{IEEEkeywords}
		Model order reduction, data-driven techniques, simulation functions, interface functions, correctness guarantees
	\end{IEEEkeywords}
	
	\section{Introduction}\phantomsection \label{Sec: Introduction}
	\subsection{Motivation}
	\IEEEPARstart{O}{VER} the past two decades, the utilization of formal methods to synthesize controllers for dynamical systems has gained increasing attention. This endeavor aims to formally design a controller, often endowed with a state-feedback-like architecture, to ensure adherence of the dynamical system to high-level logical specifications, including those expressed through linear temporal logic (LTL) formulas~\cite{pnueli1977temporal, baier2008principles}. Emerging applications, however, are typically high-dimensional due to the increasing complexity of modern engineering systems, the demand for advanced autonomous decision-making, and the integration of sophisticated control and optimization techniques. Regrettably, high dimensionality poses inherent difficulties to the formal synthesis of controllers, stemming from computational challenges. This challenge has motivated the control community to develop scalable methodologies.
	
	One promising approach in mitigating computational challenges is the use of \emph{model order reduction (MOR)}. In broad terms, MOR techniques center around forming simplified, low-dimensional models capable of retaining the main features of original, high-dimensional systems. Once a reduced-order model (ROM) is constructed, it can effectively replace the original system, thereby simplifying the control synthesis process for high-dimensional dynamical systems and alleviating computational complexity~\cite{antoulas2005approximation}. In particular, the results derived from a ROM can be systematically transferred to its original system by designing an interface function that establishes a precise correspondence between the controllers of both systems. By formally characterizing the closeness between the output trajectories of the two systems via the notion of simulation functions (SFs), one can guarantee that the original system satisfies a high-level logic specification analogous to its ROM, albeit subject to guaranteed error bounds~\cite{lavaei2022automated}.
	
	While MOR techniques provide effective means for formally synthesizing controllers for high-dimensional systems, the existing literature typically relies on the availability of a \emph{precise model} of the original system. However, such models in applications are often  unavailable or too complex to be of any use. To address this issue, the corresponding literature suggests two distinct perspectives. The first approach is to tackle the problem employing \emph{indirect} data-driven methods. Within this framework, the data gathered from high-dimensional systems is utilized to derive models using system identification techniques, followed by leveraging model-based approaches to construct the associated ROMs. Nevertheless, even if identifying a high-dimensional system and obtaining a valid model is feasible, model-based techniques can still pose computational challenges for constructing ROMs, particularly for systems with complex nonlinear dynamics~\cite{kerschen2006past,hou2013model}. On the contrary, the second perspective, to which our framework belongs, addresses the problem by adapting \emph{direct} data-driven approaches, in which the intermediate step of identifying the system is circumvented, and the ROMs can be constructed in a single step~\cite{dorfler2022bridging}.
	
	\subsection{Related Studies}
	In the field of systems and control, MOR methods can be roughly categorized into three main categories: \emph{(i)} energy-based approaches, \emph{(ii)} Krylov methods, and \emph{(iii)} SF-based methodologies.
	The general idea behind energy-based approaches, including balanced truncation~\cite{moore1981principal, enns1984model, benner2011lyapunov, prajna2005model, sandberg2010extension, besselink2014model, borja2021extended, gugercin2004survey} and Hankel-norm methods~\cite{glover1984all, safonov1990optimal, kawano2016model}, is to reduce the system's dimensions by truncating the state variables that either have no impact on the system's input--output behavior or whose influence is negligible. As a result, the error between the input--output behavior of the original system and that of the ROM remains small.
	Krylov methods, which depend on interpolation and/or the regulations of moment matching, reside in the second category~\cite{feldmann1995efficient, grimme1997krylov, gallivan2004sylvester, gallivan2004model, astolfi2007new, astolfi2010model, simard2023moment, shakib2023time, moreschini2024closed, doebeli2024polynomial}. These methods aim to construct ROMs that approximate the system's input--output behavior with a desired level of accuracy by matching a finite number of moments, typically derived from the system's response, around selected expansion points. This ensures accurate approximation in specific regions of interest (\emph{e.g.,} near an operating point).
	Finally, the third category, to which this paper belongs, centers around generating ROMs by leveraging SFs to establish a similarity relation between the original high-dimensional systems and their corresponding ROMs~\cite{girard2009hierarchical,zamani2017compositional, lavaei2017compositional, arcak2016networks, lavaei2020compositional, zhong2023automata, lavaei2019compositional}. It is worth noting that MOR techniques have also been adopted for interconnected networks in the literature~\cite{ishizaki2014dissipativity, besselink2015clustering, ishizaki2013structured, besselink2014model1}.
	
	\BS{Balanced truncation and related empirical Gramian methods are commonly classified as energy-based MOR approaches, as controllability and observability Gramians are obtained from dual Lyapunov equations, while their empirical counterparts approximate Gramian-related quantities through simulations~\cite{himpe2021comparing}. For nonlinear systems, extensions based on energy functionals as well as differential and empirical generalizations have also been developed. Conceptually, the SF-based framework can likewise be viewed as energy-based, since it employs a Lyapunov-like function satisfying a dissipation inequality. The key distinction, however, lies in the notion of error: balancing-based methods apply the same input to both the full-order and reduced-order models and quantify the reduction error as the discrepancy between their input–output maps, with classical balanced truncation providing system-norm bounds for linear systems (\emph{e.g.}, $\mathcal{H}_\infty$)~\cite{himpe2021comparing}. In contrast, SF-based frameworks permit different inputs, coupled through an interface, while delivering explicit \emph{a priori} trajectory-wise bounds on the output mismatch, thereby enabling their use in specification-oriented controller synthesis. Nevertheless, it is worth noting that both energy-based and Krylov-based methods may exhibit more favorable scalability properties compared to SF-based approaches.}
	Despite their differences, the above-mentioned studies have been conducted under the common assumption of the availability of the \emph{exact} mathematical model of the system, which is often not the case in practical scenarios.
	
	In the first two categories above, several studies have been conducted over the past few years that utilized data from the system rather than assuming the availability of the exact system's model.  In this regard, data-driven balanced truncation methods based on persistently exciting data are presented in~\cite{rapisarda2011identification} and~\cite{markovsky2005algorithms}. In~\cite{gosea2022data}, a reformulation of balanced truncation is introduced, the central theme of which is the estimation of Gramian-related quantities using convergent numerical quadratures. In~\cite{burohman2023data}, a data-driven MOR framework is presented that operates on noisy data, assuming a known noise model.
	As for nonlinear systems, empirical balanced truncation is investigated in studies such as~\cite{lall2002subspace} and~\cite{kawano2021empirical}, while data-driven MOR for monotone nonlinear systems is explored in~\cite{kawano2019data}.
	
	Within the second category, \cite{scarciotti2017data} proposes a data-driven framework for MOR via moment matching, utilizing time-domain measurement data. This approach commences by computing the system moments from the time-domain observables, after which ROMs are constructed. By leveraging the concept of swapped interconnection, \cite{mao2022data} presents an algorithm that asymptotically estimates an arbitrary number of moments of a linear system using a single time-domain sample. Additionally, \cite{mao2024data} investigates an MOR problem where the goal is to simultaneously satisfy a set of linear and nonlinear interpolation conditions. Finally, the recent work~\cite{bhattacharjee2025signal} proposes a class of ROMs constructed from input--output data that achieve moment matching even when both the underlying system and the signal generator are unknown.
	
	While the aforementioned studies in the first two categories are highly promising, they are mainly designed for stability analysis and input--output behaviors. Thus, even though valuable, these approaches cannot be utilized to ensure the satisfaction of complex LTL properties such as safety, reachability, and reach-while-avoid. This limitation motivates the development of data-driven MOR techniques within the third category.
	
	\subsection{Original Contributions}
The main objective of this work is to develop a data-driven approach for constructing ROMs of input-affine nonlinear systems with unknown dynamics, thereby facilitating control design to satisfy complex specifications. The primary contributions of the paper are stated in the sequel:
	\begin{enumerate}
		\item[\textit{(i)}] \BS{Building upon the notion of SFs, specifically adapted in our framework to account for noise in the collected data as well as other sources of uncertainty, we establish guarantees that the output trajectories of the original and reduced-order systems remain within a prescribed bound (Theorem~\ref{thm: cont: model-based}).}
		\item[\textit{(ii)}] \BS{By imposing a specific structure on the interface functions, designed to steer the system toward satisfying complex specifications, we derive a parameterization in terms of the system’s unknown matrices, which serves as the basis for the subsequent developments (Lemma~\ref{lemma: ct_data_closed}).}
		\item[\textit{(iii)}] \BS{We propose data-dependent conditions in the form of a semidefinite program (SDP) for constructing both an SF and the corresponding interface function, without system identification (Theorem~\ref{thm: main-cont}). The proposed SDP relies on data collected from a single experiment, which are corrupted by unknown-but-bounded noise with a known bound. The proposed formulation also explicitly incorporates minimizing the closeness bound between the output trajectories of the two systems and yields parameters from which the bound can be computed directly.}
		\item[\textit{(iv)}] \BS{While the preceding result assumes access to a sufficiently rich library of functions, referred to as a dictionary, capable of spanning all nonlinear terms in the true system dynamics, this requirement may be restrictive in practice. To address this limitation, we extend the proposed framework to accommodate incomplete dictionaries (Corollary~\ref{proposition}). Finally, we provide a systematic pipeline for implementing the proposed approach (Algorithm~\ref{Alg:1}).}
	\end{enumerate}
	To demonstrate the effectiveness of the proposed data-driven approach, we present simulation results on two case studies with unknown and highly nonlinear dynamics, {including a circuit of chained inverter gates with $20$ state variables.}
	
	A limited subset of our results was recently presented at a conference~\cite{samari2025model}. The framework in the current paper differs from~\cite{samari2025model} in {three} fundamental aspects. Firstly, our data-driven MOR methodology is applicable to input-affine \emph{nonlinear} systems,
	whereas the framework in~\cite{samari2025model} is for \emph{linear} control systems. This leads to a significant structural modification in the design of the interface function, {while simultaneously enhancing the applicability of the proposed framework.} {Additionally, in contrast to~\cite{samari2025model}, the proposed framework explicitly accommodates a more realistic setting in which the available data are corrupted by noise. Finally, the proposed framework requires only a single set of input–state data from the system. In contrast,~\cite{samari2025model} relies on two such datasets, one of which must be collected under zero input, thereby imposing a more demanding data acquisition procedure than that required by the proposed approach.}
	
	\subsection{Notation}
	The set of real numbers is denoted by $\R$, non-negative real numbers by $\Rpz$, and positive real numbers by $\Rp$. Sets of non-negative and positive integers are represented by $\N = \{0, 1, 2, \ldots\}$ and $\Np = \{1, 2, \ldots\}$, respectively. An $n \times n$ identity matrix is denoted by $\I_n$, while the $n$-dimensional vector of ones is represented by $\One_n$. Moreover, an $n \times m$ zero matrix is denoted by $\bbzero_{n \times m}$, while $\bbzero_{n}$ stands for the zero vector in $\mathbb{R}^n$. Given $N$ vectors $x_i \in \R^n$, the matrix $x=[x_1 \, \, \ldots \,\, x_N]$ has dimensions $n \times N$. The Euclidean norm of a vector $x\in\R^{n}$ is indicated by $\BS{\vert x\vert}$, while the induced 2-norm of a matrix $Q$ is represented by $\Vert Q \Vert$. {Given a symmetric matrix $P$, the notation $P \succ 0$ ($P \succeq 0$) indicates that $P$ is positive (semi)definite, whereas $P \prec 0$ ($P \preceq 0$) denotes that $P$ is negative (semi)definite.} Given a symmetric matrix $P$, the minimum and maximum eigenvalues of $P$ are represented by $\lambda_{\min}(P)$ and $\lambda_{\max}(P)$, respectively.
	A star $(\star)$ in a symmetric matrix indicates the transposed element in the symmetric position. \BS{The rank of a matrix $P$ is denoted by $\operatorname{rank}(P)$.}
	We denote the supremum of a function $f : \Rpz \rightarrow \R^n$ by $\BS{\vert f \vert_{\infty} \coloneq (\text{ess}) \sup \{{\vert f(t) \vert}, t \geq 0\}}$. For a given matrix $B$, \BS{$\mathrm{b} \in \operatorname{col}(B)$ indicates that the vector $\mathrm{b}$ lies in the column space of $B$, \emph{i.e.,} it can be expressed as a linear combination of the columns of $B$.}
	
	\section{{Problem Formulation}}\phantomsection \label{Sec: Problem Formulation}
	\subsection{Nonlinear Control Systems}\phantomsection \label{subsec: mD}
	We begin by introducing the continuous-time, input-affine nonlinear systems considered in this work.
	
	\begin{definition}[\textbf{ct-NCS}]\phantomsection \label{def: ct-NCS}
		A continuous-time input-affine nonlinear control system (ct-NCS) is characterized by
		\begin{align}
			\CT\!: \begin{cases}
				\dot{x} = f(x) + Bu,\\
				y = x,
			\end{cases}\phantomsection \label{eq: general  ct-NCS}
		\end{align}
		where $x \in X$ and $y  \in X$ are the state and output of the system, with $X \subset \R^n$ being the \BS{compact} state (and output) set, $f : X \rightarrow \R^n$ is a continuously differentiable vector function, $B \in \R^{n \times m}$ is the control matrix, and the system's control input is denoted by $u \in U$, with $U \subset \R^m$ being its input set. The state reached at time $t\in\Rpz$, starting from an initial condition $x_0 \in X$ and driven by an input signal {$u \in \mathcal{U}$}, is denoted by $x_{x_0u}(t)$, where $\mathcal{U}$ is a subset of the set of all measurable functions of time, from open intervals in $\Rpz$ to $U$\footnote{With a slight abuse of notation, we use the symbol $u$ to denote both elements of $U$ and elements of $\mathcal{U}$. The intended meaning, whether referring to an input value or an input function, will be clear from the context; an analogous convention applies to $\hat{u}$ (cf., Definition~\ref{def: ct-ROM}).}. Likewise, the output at time $t\in\Rpz$ is denoted by $y_{x_0u}(t)$. Given that $y=x$, the state and output trajectories of the ct-NCS described in~\eqref{eq: general  ct-NCS} are identical.
	\end{definition}
	
	Without loss of generality, the ct-NCS in~\eqref{eq: general  ct-NCS} can be equivalently expressed as
	\begin{align}\phantomsection \label{eq: ct-NCS-tmp}
		\CT\!: \begin{cases}
			\dot{x} = A_{\ast}\mathcal{D}_{\ast}(x) + Bu,\\
			y = x,
		\end{cases}
	\end{align}
	where $\mathcal{D}_{\ast} : X \rightarrow \R^{d_{\ast}}$ is a continuously differentiable vector function, and $A_{\ast} \in \R^{n \times d_{\ast}}$ is the constant system matrix. In this work, the matrices $A_{\ast}$ and $B$, as well as the exact structure of $\mathcal{D}_{\ast}(x)$, are considered \emph{unknown}. However, we make the subsequent assumption regarding $\mathcal{D}_{\ast}(x)$.
	
	\begin{assumption}[\textbf{On $\mathcal{D}_{\ast}(x)$}]
		\phantomsection
		\label{assump-on-D}
		A continuously differentiable vector function $\mathcal{D} : X \rightarrow \R^{d}$, referred to as a \emph{dictionary}, is known and satisfies
		\begin{align}
			\mathcal{D}_{\ast}(x) = \Theta \mathcal{D}(x), \phantomsection \label{eq:tmpD}
		\end{align}
		for some unknown matrix $\Theta \in \R^{d_{\ast} \times d}$.
	\end{assumption}
	
	Under Assumption~\ref{assump-on-D}, the ct-NCS in~\eqref{eq: ct-NCS-tmp} can be rewritten as
	\begin{align}\phantomsection \label{eq: ct-NCS1}
		\CT\!: \begin{cases}
			\dot{x} = A\mathcal{D}(x) + Bu,\\
			y = x,
		\end{cases}
	\end{align}
	where the \emph{unknown} matrix $A \in \R^{n \times d}$ is defined as $A \coloneq A_{\ast} \Theta$, and $\mathcal{D} : X \rightarrow \R^{d}$ is the available dictionary.
	We consider that $\mathcal{D}$ is endowed with both linear and nonlinear terms, structured as
	\begin{align}\phantomsection \label{Dictionary}
		\mathcal{D}(x) = \begin{bmatrix}
			x\vspace{-.225cm}\\
			\tikz\draw [thin,dashed] (0,0) -- (1.25,0);\\
			\mathcal{N}(x)\!
		\end{bmatrix}\!\!,
	\end{align}
	where $\mathcal{N}(x) : X \rightarrow \R^{d - n}$ comprises solely \emph{nonlinear} basis functions. \BS{By defining $A \coloneq \begin{bmatrix}
			A_1 & A_2
		\end{bmatrix}$, with $A_1 \in \R^{n \times n}$ and $A_2 \in \R^{n \times (d - n)}$ being \emph{unknown}, and considering the dictionary in~\eqref{Dictionary}, one can reformulate the ct-NCS in~\eqref{eq: ct-NCS1} as
		\begin{align}\phantomsection \label{eq: ct-NCS}
			\CT\!: \begin{cases}
				\dot{x} = A_1 x + A_2 \, \mathcal{N}(x) + Bu,\\
				y = x,
			\end{cases}
		\end{align}
		which will serve as the focus of the subsequent analysis.}
	We use the tuple $\CT = (X, U, X, A, \mathcal{D}, B, \I_n)$ to denote the reformulated ct-NCS in~\eqref{eq: ct-NCS}. 
	
	We refer to $\CT$ as a system with a (partially) unknown model, given that its matrices $A$ and $B$ are unknown, while the dictionary $\mathcal{D}(x)$ is available, a scenario that closely aligns with many applications.
	
	\begin{remark}[\textbf{On Assumption~\ref{assump-on-D}}] \phantomsection \label{rem: dictionary}
		Assumption~\ref{assump-on-D} requires that the available dictionary $\mathcal{D}$ be sufficiently \emph{extensive} to capture all possible nonlinear terms that may arise in the system's dynamics, even at the expense of including superfluous terms. This assumption is not overly restrictive, as in many practical applications (\emph{e.g.}, electrical and mechanical engineering systems), the necessary information about the system's dynamics can often be derived from first principles, aligning with the basis functions in $\mathcal{D}$. The system parameters $A$ and $B$, however, may not be available. \BS{Nevertheless, in Section~\ref{subsec:incomplete_dic}, we relax Assumption~\ref{assump-on-D} by allowing some system nonlinearities to be neglected, and thus not included in the dictionary defined in~\eqref{Dictionary}.}
	\end{remark}
	
	\begin{remark}[\textbf{On the Lifting of $f(x)$ to $\mathcal{D}(x)$}]\phantomsection\label{rem:koopman}
		\BS{Assumption~\ref{assump-on-D} enables the reformulation of the ct-NCS in~\eqref{eq: general  ct-NCS} into the form in~\eqref{eq: ct-NCS1}, and subsequently, for notational simplicity, into~\eqref{eq: ct-NCS}. The representation in~\eqref{eq: ct-NCS1} is linear with respect to the feature vector $\mathcal{D}(x)$ (\emph{i.e.}, the dictionary defined in~\eqref{Dictionary}). This is achieved by lifting the nonlinear vector function $f(x)$ in~\eqref{eq: general  ct-NCS} into a higher-dimensional feature space defined by $\mathcal{D}(x)$ (\emph{i.e.}, from dimension $n$ to $d$, where typically $d \gg n$), thereby yielding a parameterization of $f(x)$ that is linear in the lifted features.
        From a high-level perspective, this procedure has also been investigated in the literature, particularly through approaches grounded in Koopman operator theory~\cite{10565947,9516947} and immersion-based techniques~\cite{wang2020data}.
        Moreover, the proposed framework minimizes the effect of nonlinearities (cf., Section~\ref{Subsec:Disc1}),
        suggesting a potential connection to~\cite{10565947,wang2020data}, where the objective is to design control laws that render the closed-loop system linear; see also~\cite{10565947} for a recent discussion on the interplay between feedback linearization and Koopman operator theory. We emphasize, however, that feedback linearization is not the focus of the present work.}
	\end{remark}
	
	\subsection{Reduced-Order Models}
     We aim to construct a linear ROM for the ct-NCS~\eqref{eq: ct-NCS} using data, as synthesizing controllers for linear systems to satisfy complex specifications is considerably more tractable than synthesizing controllers for nonlinear systems. Having motivated the adoption of linear ROMs, we now formally introduce them in the following definition.
	
	\begin{definition}[\textbf{ct-ROM}]
		\phantomsection \label{def: ct-ROM}
		A continuous-time ROM (ct-ROM) of the ct-NCS in~\eqref{eq: ct-NCS} is characterized by
		\begin{align}
			\CTROM\!: \begin{cases}
				\dot{\hat{x}} = \hat{A}\hat{x} + \hat{B}\hat{u},\\
				\hat{y} = \hat{C}\hat{x},
			\end{cases}\phantomsection \label{eq: ct-ROM}
		\end{align}
		where $\hat{A} \in \R^{\hat{n} \times \hat{n}}$, $\hat{B} \in \R^{\hat{n} \times \hat{m}}$, and $\hat{C} \in \R^{n \times \hat{n}}$ are the matrices of the ct-ROM, with potentially $\hat{n} \ll n$. The \BS{compact} state, input, and output sets of the ct-ROM are denoted by $\hat{X} \subset \R^{\hat{n}}$, $\hat{U} \subset \R^{\hat{m}}$, and $\hat{Y} \subset \R^{n}$, respectively, where $\hat{x} \in \hat{X}$, $\hat{u} \in \hat{U}$, and $\hat{y} \in \hat{Y}$. 
		The state of the ct-ROM at time $t\in\Rpz$, starting from an initial condition $\hat{x}_0 \in \hat{X}$ and subject to an input signal {$\hat{u} \in \hat{\mathcal{U}}$}, is denoted by $\hat{x}_{\hat{x}_0\hat{u}}(t)$, where $\hat{\mathcal{U}}$ is a subset of the set of all measurable functions of time, from open intervals in $\Rpz$ to $\hat{U}$. The corresponding output at time $t$, denoted by $\hat{y}_{\hat{x}_0\hat{u}}(t)$, is defined by $\hat{y}_{\hat{x}_0\hat{u}}(t) = \hat{C} \hat{x}_{\hat{x}_0\hat{u}}(t)$. The tuple $\CTROM = (\hat{X}, \hat{U}, \hat{Y}, \hat{A}, \hat{B}, \hat{C})$ is used to represent the ct-ROM in~\eqref{eq: ct-ROM}. 
	\end{definition}
	
	Having presented the ct-NCS $\CT$ and its corresponding ct-ROM $\CTROM$, we now introduce the notion of SFs, upon which the closeness between the output trajectories of the two systems can be formally quantified.
	
	\subsection{Simulation Functions}
	SFs are defined over the Cartesian product of the state spaces to quantify the proximity between the output trajectories of $\CT$ and $\CTROM$. This concept ensures that the discrepancy between the output trajectories of the two systems is constrained within a quantified error bound. Below, we provide the definition of SFs, adopted with some modifications from~\cite{zamani2017compositional}.
	
	\begin{definition}[\textbf{SF}]
		\phantomsection
		\label{def: SF-continuous}
		Consider
		$\CT = (X, U, X, A, \mathcal{D}, B, \I_n)$ and
		$\CTROM = (\hat{X}, \hat{U}, \hat{Y}, \hat{A}, \hat{B}, \hat{C})$.
		A function $\CTSF : X \times \hat{X}\to\Rpz$ is
		called an SF from $\CTROM$  to $\CT$ if there exist constants $\alpha,\kappa, \rho\in \R^+$ \BS{and $\eta \in \Rpz$} such that
		\begin{subequations}
			\begin{itemize}
				\item $\forall x\!\in\! X,\forall \hat x\!\in\!\hat{X},$
				\begin{align}
					\alpha\vert x-\hat{C}\hat x\vert^2\le  \CTSF(x,\hat x),\phantomsection \label{eq: con1-def-cont}
				\end{align}
				\item $\forall x\in X,\forall\hat x\in\hat{X}, \forall \hat u \in \hat{U}, \exists u \in U,$ such that
				\begin{align}
					&\mathscr{L} \CTSF(x,\hat x) \leq - \kappa \CTSF(x,\hat x) + \rho \,\vert \hat u \vert^2 \, \BS{+ \, \eta}, \phantomsection \label{eq: con2-def-cont}
				\end{align}
			\end{itemize}
		\end{subequations}
		where $\mathscr{L} \CTSF$ is the Lie derivative of $\CTSF$
		with respect to the dynamics in~\eqref{eq: ct-NCS} and \eqref{eq: ct-ROM}, defined as 
		\begin{align}
			\mathscr{L}\CTSF(x,\hat x)  & =   \partial_x \CTSF(x,\hat x)(\BS{A_1 x + A_2 \, \mathcal{N}(x)} + Bu) \notag\\ &\hspace{0.4cm}+ \partial_{\hat x}\CTSF(x,\hat x)(\hat{A}\hat{x} + \hat{B}\hat{u}),\phantomsection \label{eq: Lie derivative}
		\end{align}
		where $\partial_x \CTSF(x, \hat x) = \frac{\partial\CTSF(x, \hat x)}{\partial x}$ and $\partial_{\hat x} \CTSF(x, \hat x) = \frac{\partial \CTSF(x, \hat x)}{\partial \hat x}$.
	\end{definition}
	
	An SF intuitively implies that if the initial outputs of \(\CT\) and \(\CTROM\) are close (as ensured by condition~\eqref{eq: con1-def-cont}), they will remain close over time (as ensured by condition~\eqref{eq: con2-def-cont})~\cite{tabuada2009verification}.
	
	\begin{remark}[\textbf{Interface Function}]\phantomsection \label{rem: interface-cont}
		Condition~\eqref{eq: con2-def-cont} essentially indicates the existence of a function $u = u_{\hat{u}}(x, \hat{x}, \hat{u})$ that satisfies this condition. This function, referred to as the \emph{interface function}, establishes a bridge between the control inputs $u$ and $\hat{u}$, facilitating the refinement of a synthesized control input $\hat{u}$ for $\CTROM$ into a control input $u$ for $\CT$. In Section~\ref{Sec: data-continuous}, we formally design such an interface function using data, which constitutes one of the key contributions of our work.
	\end{remark}
	
	The following theorem highlights the importance of SFs by measuring the closeness between the output trajectories of a ct-NCS $\CT$ and its ct-ROM $\CTROM$.
	
	\begin{theorem}[\textbf{Closeness Guarantee}]\phantomsection \label{thm: cont: model-based}
		Consider a ct-NCS $\CT = (X, U, X, A, \mathcal{D}, B, \I_n)$ and its ct-ROM $\CTROM = (\hat{X}, \hat{U}, \hat{Y}, \hat{A}, \hat{B}, \hat{C})$ {as in Definitions~\ref{def: ct-NCS} and~\ref{def: ct-ROM}, respectively.} Suppose $\CTSF$ is an SF from $\CTROM$ to $\CT$ according to Definition~\ref{def: SF-continuous}. Then, for any \BS{$x_0 \in X$}, \BS{$\hat{x}_0 \in \hat{X}$}, and {$\hat{u} \in \hat{\mathcal{U}}$}, there exists a corresponding {$u \in \mathcal{U}$} {such that the following inequality holds for any $t \in \Rpz$:}
		\begin{align}
			\vert y_{x_0u}(t) - \hat y_{\hat x_0\hat u}(t) \vert &=  \vert x_{x_0u}(t) - \hat C \hat x_{\hat x_0\hat u}(t) \vert \notag \\ &   \leq \BS{\sqrt{\frac{ \CTSF(x_0, \hat x_0)}{\alpha} + \frac{\rho \vert \hat u \vert_{\infty}^2 + \eta}{\alpha \kappa}}.}\phantomsection \label{eq: error-cont} 
		\end{align}
	\end{theorem}\vspace{0.1cm}
	
	\begin{proof}
		\BS{First, with a slight abuse of notation, we denote $\CTSF(x(t), \hat{x}(t))$ by $\CTSF(t)$ throughout the proof; in particular, $\CTSF(0)$ refers to $\CTSF(x_0, \hat{x}_0)$. In addition, \(\dot{\CTSF}(t) = \mathscr{L} \CTSF(x(t),\hat x(t))\), as in~\eqref{eq: con2-def-cont}.
		Accordingly, for any $t \in \Rpz$, we have
		\begin{align*}
			\dot{\CTSF}(t) \leq - \kappa \CTSF(t) + \rho \vert \hat u(t) \vert^2 + \eta \leq - \kappa \CTSF(t) + \rho \vert \hat u \vert_{\infty}^2 + \eta.
		\end{align*}
    Multiplying both sides by $\mathrm{e}^{\kappa t}$ gives
		\begin{align*}
				&\mathrm{e}^{\kappa t} \dot{\CTSF}(t) \leq - \kappa \mathrm{e}^{\kappa t} \CTSF(t) + \mathrm{e}^{\kappa t} (\rho \vert \hat u \vert_{\infty}^2 + \eta)\\
				&\Longleftrightarrow \frac{\mathrm{d}}{\mathrm{d}t} (\mathrm{e}^{\kappa t} \CTSF(t)) \leq  \mathrm{e}^{\kappa t} (\rho \vert \hat u \vert_{\infty}^2 + \eta).
		\end{align*}
		Now, by integrating from $0$ to $t$, one has
		\begin{align*}
			&\mathrm{e}^{\kappa t} \CTSF(t) - \CTSF(0) \leq (\rho \vert \hat u \vert_{\infty}^2 + \eta)  \int_{0}^{t} \mathrm{e}^{\kappa s} \mathrm{d}s\\
			&\Rightarrow \CTSF(t) \leq \mathrm{e}^{-\kappa t} \CTSF(0) + (\rho \vert \hat u \vert_{\infty}^2 + \eta)  \int_{0}^{t} \mathrm{e}^{-\kappa (t - s)} \mathrm{d}s\\
			&\Rightarrow \CTSF(t) \leq \mathrm{e}^{-\kappa t} \CTSF(0) + \frac{\rho \vert \hat u \vert_{\infty}^2 + \eta}{\kappa} (1 - \mathrm{e}^{-\kappa t})\\
			&\hspace{1.21cm}\leq \CTSF(0) + \frac{\rho \vert \hat u \vert_{\infty}^2 + \eta}{\kappa}.
		\end{align*}
		The last inequality is valid since $\kappa$ is positive and $\mathrm{e}^{-\kappa t} \leq 1$ for all $t \in \Rpz$. At the same time, according to~\eqref{eq: con1-def-cont}, we have}
		\begin{align*}
			\BS{\alpha\vert x_{x_0u}(t) - \hat C \hat x_{\hat x_0\hat u}(t)\vert^2 = \alpha \vert y_{x_0u}(t) - \hat y_{\hat x_0\hat u}(t) \vert^2 \leq  \CTSF(t).}
		\end{align*}
		\BS{Hence, one has
		\begin{align*}
			&\alpha \vert y_{x_0u}(t) - \hat y_{\hat x_0\hat u}(t) \vert^2 \leq \CTSF(0) + \frac{\rho \vert \hat u \vert_{\infty}^2 + \eta}{\kappa}\\
			&\Rightarrow \vert y_{x_0u}(t) - \hat y_{\hat x_0\hat u}(t) \vert \leq \sqrt{\frac{ \CTSF(0)}{\alpha} + \frac{\rho \vert \hat u \vert_{\infty}^2 + \eta}{\alpha \kappa}}\\
			&\hspace{3.3cm}= \sqrt{\frac{ \CTSF(x_0, \hat x_0)}{\alpha} + \frac{\rho \vert \hat u \vert_{\infty}^2 + \eta}{\alpha \kappa}},
		\end{align*}
		thereby concluding the proof.}
	\end{proof}
	
	\begin{remark}[\textbf{On Theorem~\ref{thm: cont: model-based}}] \phantomsection \label{rem: eta_and_error}
		\BS{While the inclusion of the term $\eta$ in condition~\eqref{eq: con2-def-cont} enables a formal characterization of the effect of noise in the collected data, as well as other sources of uncertainty, it inevitably increases the mismatch between the output trajectories of the two systems. Consequently, minimizing $\eta$ constitutes a central design objective. Similarly, obtaining a tighter closeness guarantee necessitates minimizing $\rho$, which forms an additional design objective. Both objectives are explicitly addressed in Theorem~\ref{thm: main-cont} and Corollary~\ref{proposition}. We finally note that if $\CTSF(x_0, \hat{x}_0) = 0$, the first term in~\eqref{eq: error-cont} vanishes, thereby further tightening the closeness guarantee.}
	\end{remark}
	
	\begin{remark}[\textbf{Complex Specifications Enforcement}]\phantomsection \label{rem: spec}
		Since Theorem~\ref{thm: cont: model-based} provides a  closeness guarantee between the output trajectories of $\CT$ and $\CTROM$, the proposed results can be utilized to enforce a variety of complex properties beyond stability, such as \emph{safety, reachability, and reach-while-avoid}~\cite{tabuada2009verification}. In particular, as desired properties are typically defined over the output space of dynamical systems, a ct-ROM is advantageous for enforcing these properties over simplified lower-dimensional systems; the results can then be refined back to the complex original nonlinear system via an interface function, while quantifying a guaranteed error bound for their closeness as offered in~\eqref{eq: error-cont}.
	\end{remark}
	
	To develop an SF and measure the proximity between $\CT$ and $\CTROM$, as described in Theorem~\ref{thm: cont: model-based}, it becomes clear that matrices $A$ and $B$ must be known due to their role in~\eqref{eq: Lie derivative}. Acknowledging this key challenge, we now formally state the main problem addressed in the paper.
	
	\begin{resp}
		\begin{problem}\phantomsection \label{problem 1}
			Consider a ct-NCS~\eqref{eq: ct-NCS}, where matrices $A$ and $B$ are unknown, while the extended dictionary $\mathcal{D}(x)$ is available. Develop a data-driven framework that is guaranteed to construct a ct-ROM for the ct-NCS, as described in~\eqref{eq: ct-ROM}. Additionally, design an interface function $u = u_{\hat u}(x, \hat x, \hat u)$, while constructing an SF from ct-ROM to ct-NCS, as defined in Definition~\ref{def: SF-continuous}.
		\end{problem}
	\end{resp}
	
	\section{\BS{Data-Driven Framework}}\phantomsection \label{Sec: data-continuous}
	\BS{In this section, we present the proposed data-driven framework for addressing Problem~\ref{problem 1}. Specifically, Section~\ref{subsec:data_collection} details the data collection procedure and outlines the assumptions on the collected data. Based on these foundations, Section~\ref{subsec:dd-ctrom-sfs} develops the main data-driven results. Finally, Section~\ref{subsec:incomplete_dic} extends the framework to scenarios in which the dictionary in~\eqref{Dictionary} is incomplete and does not encompass all nonlinear terms of the system dynamics.}
	\subsection{{Data Collection and Underlying Assumptions}}\phantomsection \label{subsec:data_collection}
	\BS{The first step in our approach is to gather data from the system. To this end, under the key assumption of direct measurability of all state variables (Definition~\ref{def: ct-NCS}),} we collect data from the unknown ct-NCS $\CT$ over the time interval $[t_0, t_0 + (T - 1)\tau]$, where $T \in \Np$ is the number of collected samples, and $\tau \in \Rp$ is the sampling time:
	\begin{subequations}\phantomsection \label{eq: data_ct} 
		\begin{align}
			\UC &= \begin{bmatrix}
				u(t_0) & u(t_0 + \tau) & \dots & u(t_0 + (T - 1)\tau)
			\end{bmatrix}\!\!,\phantomsection \label{eq: UC}\\
			\XC &= \begin{bmatrix}
				x(t_0) & x(t_0 + \tau) & \dots & x(t_0 + (T - 1)\tau)
			\end{bmatrix}\!\!,\phantomsection \label{eq: XC}\\
			\BS{\XdC} &= \begin{bmatrix}
				\dot x(t_0) & \dot x(t_0 + \tau) & \dots & \dot x(t_0 + (T - 1)\tau)
			\end{bmatrix}\!\!.\phantomsection \label{eq: XdC}
		\end{align}
	\end{subequations}
	\BS{Based on the data in~\eqref{eq: XC} and $\mathcal{N}(x)$ in~\eqref{Dictionary}, we now construct the data matrix
	\begin{align*}
		\pmb{\mathscr{N}} = \begin{bmatrix}
				\mathcal{N}(x(t_0)\!) & 	\!\!\mathcal{N}(x(t_0 + \tau)\!) & \!\!\dots & 	\!\!\mathcal{N}(x(t_0 + (T - 1)\tau)\!)
			\end{bmatrix} \!\!,
	\end{align*}
	followed by forming the data matrix $\DC = \begin{bmatrix}
			\XC^\top &  \pmb{\mathscr{N}}^\top
		\end{bmatrix}^{\! \! \top}$.}
	\BS{While many studies assume the exact availability of $\XdC$ in~\eqref{eq: XdC} (\emph{e.g.},~\cite{samari2025model}), in practice, $\XdC$ is inherently imprecise and cannot be measured directly. Hence, we approximate its elements as
		$\dot{x}_i\left(t_0+k\tau\right)=\frac{x_i\left(t_{0}+(k+1)\tau\right)-x_i\left(t_0+k\tau\right)}{\tau}+e_i\left(t_0+k\tau\right)$, with $i\in\{1, \ldots, n\}$, and $ k \in \{0, 1, \dots, T-1\}$,
		where the approximation error $e_i\left(t_0+k\tau\right)$ is proportional to $\tau$ and can be considered as noise\footnote{\BS{When approximating $\dot{x}_i(t_0+k\tau)$ for $k\in\{0,1,\ldots,T-1\}$ via forward differences, one additionally requires $x(t_0+T\tau)$ and, consequently, $u(t_0+T\tau)$. These extra samples are solely used to compute $\dot{x}(t_0+(T-1)\tau)$ and are not included in the data matrices $\XC$ and $\UC$.}}.
		Thus, rather than the exact $\XdC$ in~\eqref{eq: XdC}, which is not available, we have access to $\XdCm$ (\emph{i.e.}, $\XdCm$ is the approximated state-derivative data matrix), where $\XdC \coloneq \XdCm + \EC$, with
		\begin{align*}
			\EC = \begin{bmatrix}
				e(t_0) & e(t_0 + \tau) & \dots & e(t_0 + (T - 1)\tau)
			\end{bmatrix}
		\end{align*}
		capturing the approximation error. While $\EC$ is itself unknown, we impose the following assumption on it.}
	
	\begin{assumption}[\textbf{On $\EC$}] \phantomsection \label{assump:noise}
		\BS{The unknown data matrix $\EC$ satisfies
		\begin{align}
			\EC \EC^\top \preceq \mathcal{E} \mathcal{E}^\top\!\!, \phantomsection \label{eq: noise_bound_matrix}
		\end{align}
		for some known $\mathcal{E}$ of appropriate dimensions, implying that over the finite data collection interval, the energy of the noise remains bounded~\cite{van2020noisy}.}
	\end{assumption}
	
	\begin{remark}[\textbf{On Assumption~\ref{assump:noise}}] \phantomsection \label{rem: Assump-noise}
		\BS{In practice, a representative scenario in which Assumption~\ref{assump:noise} holds arises when a known constant $\varphi_1$ satisfies $|e_j|^2 \leq \varphi_1$ for all $j \in \{1, \ldots, T\}$, where $e_j$ denotes the $j$-th column of $\EC$. In such a case, by taking an arbitrary $\boldsymbol{\mathrm{z}} \in \R^{n}$, and by recalling the Cauchy–Schwarz inequality~\cite{bhatia1995cauchy}, we have
		\begin{align*}
			\boldsymbol{\mathrm{z}}^{\top} \EC \EC^{\top}\boldsymbol{\mathrm{z}} & =\sum_{j=1}^{ T}\bigl(e_j^{\top}\boldsymbol{\mathrm{z}}\bigr)^{2} {\le}\sum_{j=1}^{ T}\vert e_j \vert ^{2}\,\vert \boldsymbol{\mathrm{z}}\vert^{2} \\
			&\le\sum_{j=1}^{ T}\varphi_1\,\vert \boldsymbol{\mathrm{z}} \vert^{2} =\varphi_1\, T\,\vert \boldsymbol{\mathrm{z}} \vert^{2} =  \boldsymbol{\mathrm{z}}^\top   \left[\varphi_1\, T \I_{n}\right]    \boldsymbol{\mathrm{z}}.
		\end{align*}
		Since this inequality holds for every vector $ \boldsymbol{\mathrm{z}}$, one has
			\(
			\EC \EC^{\top}\preceq \varphi_1 T \I_{n},
			\)
			which implies that Assumption~\ref{assump:noise} is satisfied with \( \mathcal{E} \coloneq \sqrt{\varphi_1 T} \I_{n} \), resulting in $  \mathcal{E} \mathcal{E}^\top = \varphi_1T\I_{n}$.}
	\end{remark}
	
	\BS{We further make the following assumption on the collected data, which facilitates the feasibility analysis of the proposed framework in the discussion provided in Section~\ref{Sec:Discussion}.}
	
	\begin{assumption}[\textbf{On the Collected Data}] \phantomsection \label{assumpt:full-rank}
		\BS{The data matrix
			$\begin{bmatrix}
				\UC^\top & \XC^\top & \pmb{\mathscr{N}}^\top
			\end{bmatrix}^\top$
			has full row rank, \emph{i.e.},
		\begin{align}
			 \operatorname{rank} \left( \begin{bmatrix}
					\UC^\top & \XC^\top & \pmb{\mathscr{N}}^\top
				\end{bmatrix}^\top\right) \! =  \operatorname{rank} \left( \begin{bmatrix}
					\UC\\ \DC
				\end{bmatrix}\right) \!  = \! m \! + \! d. \phantomsection \label{eq:new_rank_con}
		\end{align}}
	\end{assumption}
	
	\begin{remark}[\textbf{On Assumption~\ref{assumpt:full-rank}}] \phantomsection \label{rem:full-data}
		\BS{Assumption~\ref{assumpt:full-rank} essentially constitutes a data richness condition~\cite{de2023learning}, ensuring that the collected data are sufficiently informative for the subsequent analysis. This assumption can be viewed as a natural generalization of the rank condition
			\(
			\operatorname{rank} \big( \!
			\begin{bmatrix}
				\UC^\top & \XC^\top
			\end{bmatrix}^{\!\top} \!\!
			\big)= m+n,
			\)
			which appears in the analysis of linear systems (see \emph{e.g.},~\cite[condition~6]{de2019formulas}) and is closely related to the notion of persistency of excitation~\cite{willems2005note}. It is worth noting that, for~\eqref{eq:new_rank_con} to be potentially satisfied, the number of samples $T$ must satisfy $T \geq m + d$, and the input signal must be persistently exciting. In practice, however, ensuring that~\eqref{eq:new_rank_con} holds may require a substantially larger value of $T$ than $m + d$. We also note that while the problem of designing a persistently exciting input $\UC$, particularly in \emph{nonlinear} settings, remains largely unresolved in the literature, in experimental settings, certain well-known input signals are commonly employed and have proven effective. A typical example is
			\(
			u(t) = \sum_{k=1}^{\mathrm{K}} \varpi_k \sin(\omega_k t + \varrho_k),
			\)
			where $\varrho_k$ denotes the phase shift, the frequencies $\omega_k$ are non-commensurate, and the amplitudes $\varpi_k$ are nonzero. Alternatively, uniformly distributed noise can also be a viable option, particularly when amplitude constraints should be respected.}
	\end{remark}
	
	\subsection{\BS{Data-Driven Construction of ct-ROMs and SFs}}\phantomsection\label{subsec:dd-ctrom-sfs}
	\BS{Having presented the details associated with data collection, we now proceed with proposing our approach. To design an SF satisfying the conditions in Definition~\ref{def: SF-continuous}, we focus on quadratic SFs of the form
		\(
		\CTSF(x,\hat{x}) = (x - \mathbfcal{R}_1 \hat{x})^\top P (x - \mathbfcal{R}_1 \hat{x}),
		\)
		where \( P \succ 0 \) and \( \mathbfcal{R}_1 \in \R^{n \times \hat{n}} \) is the top block of the \emph{reduction} matrix \( \mathbfcal{R} \in \R^{d \times \hat{n}} \), given by
	\begin{align}\phantomsection \label{Reduction}
		\mathbfcal{R} =
			\begin{bmatrix}
				\mathbfcal{R}_1 \vspace{-.225cm} \\
				\tikz\draw [thin,dashed] (0,0) -- (1.25,0); \\
				\mathbfcal{R}_2
			\end{bmatrix}\! \! ,
	\end{align}
	with \( \mathbfcal{R}_2 \in \R^{(d-n) \times \hat{n}} \). In view of the adopted structure of $\CTSF(x,\hat{x})$ and the Lie derivative in~\eqref{eq: Lie derivative}, the term $\dot{x}-\mathbfcal{R}_1\dot{\hat{x}}$ naturally appears in condition~\eqref{eq: con2-def-cont}. We therefore derive a parameterization of this term in the following lemma, while simultaneously introducing the proposed structure of the interface function~$u$.}
	
	\begin{lemma}[\textbf{Parameterization of~$\dot{x}-\mathbfcal{R}_1\dot{\hat{x}}$}]\phantomsection \label{lemma: ct_data_closed}
		Given a ct-NCS $\CT$ and its ct-ROM $\CTROM$, let
			$
			S \coloneq
			\begin{bmatrix}
				B & A_1 & A_2
			\end{bmatrix}
			$. Considering an interface function of the form
		\begin{align}
			u = \BS{\begin{bmatrix}
					KP & G
			\end{bmatrix}} (\mathcal{D}(x) - \mathbfcal{R} \hat x) + \Xi \hat x + \Psi \hat u, \phantomsection \label{eq:interface}
		\end{align}
		\BS{where $K \in \R^{m \times n}$, $P \in \R^{n \times n}$, with $P \succ 0$, $G \in \R^{m \times (d - n)}$,} $\Xi \in \R^{m \times \hat n}$, and $\Psi \in \R^{m \times \hat m}$, \BS{one has the following parameterization:
		\begin{align}
			& \dot{x}-\mathbfcal{R}_1\dot{\hat{x}} \notag\\
			&=  S \begin{bmatrix}
					KP \\ \I_{n} \\ \bbzero_{(d - n) \times n}
				\end{bmatrix} (x - \mathbfcal{R}_1 \hat x) +  \Big(S \begin{bmatrix}
					\Xi - G\mathbfcal{R}_2 \\ \mathbfcal{R}_1 \\ \bbzero_{(d - n) \times \hat n}
				\end{bmatrix} - \mathbfcal{R}_1 \hat{A}  \Big) \hat x \notag \\
			& ~~ ~ +  \Big(S \begin{bmatrix}
					\Psi \\ \bbzero_{n \times \hat m} \\ \bbzero_{(d - n) \times \hat m}
				\end{bmatrix} - \mathbfcal{R}_1 \hat{B} \Big) \hat u  + S  \begin{bmatrix}
					G \\ \bbzero_{n \times (d - n)} \\ \I_{d - n}
				\end{bmatrix}  \mathcal{N}(x) . \phantomsection \label{eq:parameterization}
		\end{align}}
	\end{lemma}
	
	\begin{proof}
		\BS{Considering the ct-NCS $\CT$ in~\eqref{eq: ct-NCS} and its ct-ROM $\CTROM$ in~\eqref{eq: ct-ROM}, together with the reduction matrix in~\eqref{Reduction} and the interface function $u$ in~\eqref{eq:interface}, one has
		\begin{align*}
			&\dot{x}-\mathbfcal{R}_1\dot{\hat{x}}\\
			& = A_1 x + A_2 \, \mathcal{N}(x) + Bu - \mathbfcal{R}_1 (\hat{A}\hat{x} + \hat{B}\hat{u})\\
			& \! \overset{\eqref{eq:interface}}{=} A_1 x + A_2 \, \mathcal{N}(x) + B \begin{bmatrix}
					KP & G
				\end{bmatrix} (\mathcal{D}(x) - \mathbfcal{R}\hat x) + B \Xi \hat x \\
			& ~~~ + B \Psi \hat u - \mathbfcal{R}_1 \hat{A}\hat{x} - \mathbfcal{R}_1 \hat{B}\hat{u}\\
			& \overset{\eqref{Dictionary}}{=} A_1 x + A_2 \, \mathcal{N}(x) + B KP (x - \mathbfcal{R}_1 \hat x) + BG(\mathcal{N}(x) - \mathbfcal{R}_2 \hat x)\\
			& ~~~ + B \Xi \hat x + B \Psi \hat u - \mathbfcal{R}_1 \hat{A}\hat{x} - \mathbfcal{R}_1 \hat{B}\hat{u}.
		\end{align*}
		Now, by sorting the right-hand side and including the term $A_1 \mathbfcal{R}_1 \hat x$ through addition and subtraction, we have
		\begin{align*}
			&\dot{x}-\mathbfcal{R}_1\dot{\hat{x}}\\
			& = A_1 (x - \mathbfcal{R}_1 \hat x) + B KP (x - \mathbfcal{R}_1 \hat x) \! + \! \big ( A_1 \mathbfcal{R}_1 \! + \! B \Xi \! - \!  \mathbfcal{R}_1 \hat{A} \\
			& ~~~ -  BG\mathbfcal{R}_2 \big ) \hat x + (B\Psi - \mathbfcal{R}_1 \hat B) \hat u + (A_2 + BG)\mathcal{N}(x)\\
			& = \begin{bmatrix}
					B & A_1 & A_2
				\end{bmatrix} \begin{bmatrix}
					KP \\ \I_{n} \\ \bbzero_{(d - n) \times n}
				\end{bmatrix} (x - \mathbfcal{R}_1 \hat x)\\
			& ~~~ + \Big ( \! \begin{bmatrix}
					B & A_1 & A_2
				\end{bmatrix} \begin{bmatrix}
					\Xi - G\mathbfcal{R}_2 \\ \mathbfcal{R}_1 \\ \bbzero_{(d - n) \times \hat n}
				\end{bmatrix} - \mathbfcal{R}_1 \hat{A} \Big ) \hat x\\
			& ~~~ + \Big ( \! \begin{bmatrix}
					B & A_1 & A_2
				\end{bmatrix} \begin{bmatrix}
					\Psi \\ \bbzero_{n \times \hat m} \\ \bbzero_{(d - n) \times \hat m}
				\end{bmatrix} - \mathbfcal{R}_1 \hat{B} \Big ) \hat u\\
			& ~~~ + \begin{bmatrix}
					B & A_1 & A_2
				\end{bmatrix} \begin{bmatrix}
					G \\ \bbzero_{n \times (d - n)} \\ \I_{d - n}
				\end{bmatrix}  \mathcal{N}(x),
		\end{align*}
		leading to~\eqref{eq:parameterization} with $
		S =
		\begin{bmatrix}
			B & A_1 & A_2
		\end{bmatrix}
		$, thereby concluding the proof.}
	\end{proof}
	
	Having offered the {interface function in~\eqref{eq:interface} and the parameterization~\eqref{eq:parameterization} of the term $\dot{x}-\mathbfcal{R}_1\dot{\hat{x}}$}, we now propose the following theorem, which enables the simultaneous construction of the interface function~\eqref{eq:interface} and the SF $\CTSF(x,\hat{x})$  \BS{directly from noise-corrupted data} collected from the ct-NCS $\CT$, without identifying the ct-NCS $\CT$.
	
	\begin{theorem}[\textbf{Data-Driven Design}]\phantomsection \label{thm: main-cont}
		\BS{Given the ct-NCS $\CT$ in~\eqref{eq: ct-NCS} and its ct-ROM $\CTROM$ in~\eqref{eq: ct-ROM}, with $\hat{C} = \mathbfcal{R}_1$, let Assumptions~\ref{assump-on-D}--\ref{assumpt:full-rank} hold. Then, if there exist decision variables $Q_1 \in \R^{T \times (d - n)}$, $Q_2 \in \R^{T \times \hat n}$, $Q_3 \in \R^{T \times \hat m}$, $\Pi \in \R^{n \times n}$, with $\Pi \succ 0$, $K \in \R^{m \times n}$, $\mu_1, \, \mu_2, \, \mu_3, \, \mu_4, \, \mu_5 \in \Rp$, and $\mu_6 \in \Rpz$ such that, for some $\kappa, \, \bar{\mu}, \, \varepsilon \in \Rp$, $\hat A$, and $\hat B$, the SDP
		\begin{mini!}|s|[2]<b>
			{\substack{\mu_1, \dots, \mu_6,\\ Q_1, Q_2, Q_3,\\ K, \Pi}}{\sum_{i=1}^{3} \! \Vert Q_i \Vert \! - \!\! \sum_{j=1}^{5} \! \mu_j  \! + \! \Vert \XdCm Q_1 \Vert \! + \! \Vert \XdCm Q_3 \! - \! \XC Q_2 \hat B \Vert}
			{\label{eq: SDP}}{\label{eq: mini_SDP}}
			\addConstraint{\mu_j \leq \bar{\mu}, \quad \forall j \in \{1, \ldots, 5\},}{ \label{eq: SDP_con1} }
			\addConstraint{	\One_T^{\! \top} \, Q_2 \, \One_{\hat n} \geq \varepsilon,}{ \label{eq: SDP_con2} }
			\addConstraint{\hspace{0.12cm}\DC Q_1 = \begin{bmatrix}
						\bbzero_{n \times (d - n)} \\ \I_{d - n}
					\end{bmatrix} \! \! ,}{ \label{eq: SDP_con6} }
			\addConstraint{\XdCm Q_2 = \XC Q_2 \hat A,}{ \label{eq: SDP_con3} }
			\addConstraint{\hspace{0.05cm}\pmb{\mathscr{N}} Q_2 = \bbzero_{(d - n) \times \hat n},}{ \label{eq: SDP_con4} }
			\addConstraint{\hspace{0.17cm}\DC Q_3 = \bbzero_{d \times \hat m},}{ \label{eq: SDP_con5} }
			\addConstraint{\begin{bmatrix}
						Z & ~~ \begin{bmatrix}
							K\\
							\Pi\\
							\bbzero_{(d - n) \times n}
						\end{bmatrix}^\top + \mu_6 \XdCm \pmb{\mathds{H}}^\top\\
						\star & -\mu_6 \pmb{\mathds{H}}\pmb{\mathds{H}}^\top
					\end{bmatrix} \! \! \preceq 0}{ \label{eq: SDP_con7} }
		\end{mini!}
		has a solution, where $\pmb{\mathds{H}} \coloneq \begin{bmatrix}
				\UC^\top & \DC^\top
			\end{bmatrix}^{\! \top}$ and
		\begin{align*}
			Z \coloneq \kappa \Pi + \sum_{j=1}^{5}  \mu_j  \I_n - \mu_6 (\XdCm {\XdCm}^{\! \top} - \mathcal{E} \mathcal{E}^\top),
		\end{align*}
		then $\CTSF(x,\hat{x}) = (x - \mathbfcal{R}_1 \hat{x})^\top P (x - \mathbfcal{R}_1 \hat{x})$ is an SF from $\CTROM$  to $\CT$, with $\mathbfcal{R}_1 \coloneq \XC Q_2$, $P \coloneq \Pi^{-1}$, $\alpha \coloneq \lambda_{\min}(P)$, and
		\begin{subequations}\phantomsection \label{eq:rho_eta}
			\begin{align}
				\rho~ &  \coloneq  \frac{\Vert \XdCm Q_3 - \XC Q_2 \hat{B} \Vert^2}{\mu_4} + \frac{\lambda_{\max}(\mathcal{E} \mathcal{E}^\top) \Vert Q_3 \Vert^2}{\mu_5},\phantomsection\label{eq:rho1}\\
				\eta~ &  \coloneq  \Big( \frac{\Vert \XdCm Q_1 \Vert^2}{\mu_1} + \frac{\lambda_{\max}(\mathcal{E} \mathcal{E}^\top) \Vert Q_1 \Vert^2}{\mu_2} \Big) \max_{x \in X} \vert \mathcal{N}(x) \vert^2 \notag\\
				& \hspace{0.4cm} + \frac{\lambda_{\max}(\mathcal{E} \mathcal{E}^\top) \Vert Q_2 \Vert^2}{\mu_3} \max_{\hat x \in \hat X} \vert \hat x \vert^2 .\phantomsection\label{eq:eta1}
			\end{align}
		\end{subequations}
		Moreover, the corresponding interface function is given by~\eqref{eq:interface}, with $G \coloneq \UC Q_1$, $\mathbfcal{R}_2 \coloneq \pmb{\mathscr{N}} Q_2 = \bbzero_{(d - n) \times \hat n}$ (cf., constraint~\eqref{eq: SDP_con4}), $\Xi \coloneq \UC Q_2$, and $\Psi \coloneq \UC Q_3$.}
	\end{theorem}
	
	\begin{proof}
		\BS{The proof proceeds in two steps. Under the feasibility of the SDP~\eqref{eq: SDP}, the first step establishes that condition~\eqref{eq: con1-def-cont} holds, while the second, and more technically challenging, step demonstrates that condition~\eqref{eq: con2-def-cont} is satisfied. To proceed with the former, we first recall that $\hat C = \mathbfcal{R}_1$. Therefore, one has
			\[
			\vert x - \hat C \hat x\vert^2 = \vert x - \mathbfcal{R}_1 \hat x\vert^2.
			\]
			Given that
			\[
			\lambda_{\min}(P) \vert x - \mathbfcal{R}_1 \hat x\vert^2 \leq (x- \mathbfcal{R}_1\hat x)^\top P (x- \mathbfcal{R}_1\hat x) = \CTSF(x,\hat{x}),
			\]
			one can deduce that}
		\begin{align*}
			\BS{\lambda_{\min}(P) \vert x - \mathbfcal{R}_1 \hat x\vert^2 \leq \CTSF(x,\hat{x}),}
		\end{align*}
		\BS{showing that condition~\eqref{eq: con1-def-cont} holds with $\alpha \coloneq \lambda_{\min}(P)$, thereby concluding the first step of the proof.}
		
		\BS{To proceed with the second step, considering $P \coloneq \Pi^{-1}$ and $\CTSF(x, \hat x) = (x- \mathbfcal{R}_1\hat x)^\top P (x- \mathbfcal{R}_1\hat x)$, we have
		\begin{align*}
			\mathscr{L}\CTSF(x,\hat x)  & \overset{\eqref{eq: Lie derivative}}{=} \partial_x \CTSF(x,\hat x)(A\mathcal{D}(x) \! + \! Bu) \! + \! \partial_{\hat x} \CTSF(x,\hat x)(\hat A\hat x  \! + \! \hat B\hat u)\\
			& = 2 (x - \mathbfcal{R}_1 \hat x)^\top \Pi^{-1} \underbrace{(A\mathcal{D}(x) \! + \! Bu)}_{\dot x}\\
			& ~~~ -2 (x - \mathbfcal{R}_1 \hat x)^\top \Pi^{-1} \mathbfcal{R}_1 \underbrace{(\hat A\hat x  \! + \! \hat B\hat u)}_{\dot{\hat{x}}}\\
			& = 2 (x - \mathbfcal{R}_1 \hat x)^\top \Pi^{-1} (\dot x - \mathbfcal{R}_1 \dot{\hat{x}}).
		\end{align*}
		By virtue of Lemma~\ref{lemma: ct_data_closed}, we know that the term $\dot{x}-\mathbfcal{R}_1\dot{\hat{x}}$ can be parameterized as in~\eqref{eq:parameterization}. Hence, one has
		\begin{align}
			& \mathscr{L}\CTSF(x,\hat x) \notag\\
			& = 2 (x - \mathbfcal{R}_1 \hat x)^\top \Pi^{-1} S \begin{bmatrix}
					K\Pi^{-1} \\ \I_{n} \\ \bbzero_{(d - n) \times n}
				\end{bmatrix} (x - \mathbfcal{R}_1 \hat x)\notag\\
			& ~~~ + 2 (x - \mathbfcal{R}_1 \hat x)^\top \Pi^{-1} \overbrace{\Big (S \begin{bmatrix}
						\Xi - G\mathbfcal{R}_2 \\ \mathbfcal{R}_1 \\ \bbzero_{(d - n) \times \hat n}
					\end{bmatrix} - \mathbfcal{R}_1 \hat{A} \Big)}^{\clubsuit} \hat x \notag\\
			& ~~~ + 2 (x - \mathbfcal{R}_1 \hat x)^\top \Pi^{-1}  \overbrace{\Big(S \begin{bmatrix}
						\Psi \\ \bbzero_{n \times \hat m} \\ \bbzero_{(d - n) \times \hat m}
					\end{bmatrix} - \mathbfcal{R}_1 \hat{B} \Big)}^{\spadesuit} \hat u \notag\\
			& ~~~ + 2 (x - \mathbfcal{R}_1 \hat x)^\top \Pi^{-1} \overbrace{S  \begin{bmatrix}
						G \\ \bbzero_{n \times (d - n)} \\ \I_{d - n}
					\end{bmatrix}  \mathcal{N}(x)}^{\bigstar} \! . \phantomsection \label{eq:tmp1}
		\end{align}
		At the same time, according to the ct-NCS in~\eqref{eq: ct-NCS} and the collected data in~\eqref{eq: data_ct}, and recalling that $\XdC \coloneq \XdCm + \EC$, we have
		\begin{align}
			\XdCm  & = \overbrace{A_1 \XC + A_2 \pmb{\mathscr{N}} + B \UC}^{\XdC} - \EC \notag \\
			& = S \begin{bmatrix}
					\UC \\ \XC \\ \pmb{\mathscr{N}}
				\end{bmatrix} - \EC = S  \begin{bmatrix}
					\UC \\ \DC
				\end{bmatrix} - \EC. \phantomsection \label{eq:data_rel}
		\end{align}
		Now, we first concentrate on the term ``$\bigstar$" in~\eqref{eq:tmp1} to proceed with the proof. Considering that $G \coloneq \UC Q_1$,  the term ``$\bigstar$" can be rewritten under constraint~\eqref{eq: SDP_con6} as
		\begin{align}
			S  \begin{bmatrix}
					G \\ \bbzero_{n \times (d - n)} \\ \I_{d - n}
				\end{bmatrix} \!\!  \mathcal{N}(x) & = S \begin{bmatrix}
					\UC Q_1 \\ \DC Q_1
				\end{bmatrix} \mathcal{N}(x) = S \! \begin{bmatrix}
					\UC  \\ \DC 
				\end{bmatrix} \! Q_1 \mathcal{N}(x) \notag\\
			& \! \overset{\eqref{eq:data_rel}}{=} \! (\XdCm + \EC) Q_1 \mathcal{N}(x). \phantomsection\label{eq:tmp13}
		\end{align}
		Consequently, the fourth term in~\eqref{eq:tmp1} can be rewritten as
		\begin{align}
			& 2 (x - \mathbfcal{R}_1 \hat x)^\top \Pi^{-1} (\XdCm + \EC) Q_1 \mathcal{N}(x) \notag\\
			& \! = \! 2 (x \! - \! \mathbfcal{R}_1 \hat x)^{\! \! \top} \Pi^{-1} \XdCm Q_1 \mathcal{N}(x) \! + \! 2 (x \! - \! \mathbfcal{R}_1 \hat x)^{\! \! \top} \Pi^{-1} \EC Q_1 \mathcal{N}(x).  \phantomsection \label{eq:tmp2}
		\end{align}
		We now apply the Cauchy–Schwarz inequality~\cite{bhatia1995cauchy}, \emph{i.e.}, $a^\top b \le |a| |b|$ for any $a,b \in \R^{n}$, followed by Young’s inequality~\cite{young1912classes}, \emph{i.e.}, $|a| |b| \le \tfrac{\mu}{2}|a|^2 + \tfrac{1}{2\mu}|b|^2$ for any $\mu \in \Rp$, to both terms in~\eqref{eq:tmp2}; consequently, for arbitrary $\mu_1,\mu_2 \in \Rp$, we obtain
		\begin{align*}
			& 2 (x - \mathbfcal{R}_1 \hat x)^\top \Pi^{-1} (\XdCm + \EC) Q_1 \mathcal{N}(x)\\
			& \leq  (\mu_1 + \mu_2) (x - \mathbfcal{R}_1 \hat x)^\top \Pi^{-1} \Pi^{-1} (x - \mathbfcal{R}_1 \hat x)\\
			& ~~~ + \frac{1}{\mu_1} \Vert \XdCm Q_1 \Vert^2 \vert \mathcal{N}(x) \vert^2 + \frac{1}{\mu_2} \mathcal{N}^\top(x) Q_1^\top \EC^\top \EC Q_1 \mathcal{N}(x).
		\end{align*}
		Note that $\Pi^{-1^\top} = \Pi^{-1}$ since $\Pi \succ 0$.  According to Assumption~\ref{assump:noise}, we have
			\( \EC \EC^\top \preceq \mathcal{E} \mathcal{E}^\top, \)
			for some known $\mathcal{E}$ of appropriate dimensions, implying that
			\(
			\lambda_{\max}(\EC \EC^\top) \leq \lambda_{\max}(\mathcal{E} \mathcal{E}^\top)
			\).
			Also, since
			\(
			\lambda_{\max}(\EC \EC^\top) = \lambda_{\max}(\EC^\top \EC)
			\), one can conclude that
			\(
			\lambda_{\max}(\EC^\top \EC) \leq \lambda_{\max}(\mathcal{E} \mathcal{E}^\top)
			\). Hence, we have
		\begin{align*}
			& 2 (x - \mathbfcal{R}_1 \hat x)^\top \Pi^{-1} (\XdCm + \EC) Q_1 \mathcal{N}(x)\\
			& \leq  (\mu_1 + \mu_2) (x - \mathbfcal{R}_1 \hat x)^\top \Pi^{-1} \Pi^{-1} (x - \mathbfcal{R}_1 \hat x)\\
			& ~~~ + \frac{1}{\mu_1} \Vert \XdCm Q_1 \Vert^2 \vert \mathcal{N}(x) \vert^2 + \frac{\lambda_{\max}(\mathcal{E} \mathcal{E}^\top)}{\mu_2} \Vert Q_1 \Vert^2 \vert \mathcal{N}(x)\vert^2\\
			& \leq  (\mu_1 + \mu_2) (x - \mathbfcal{R}_1 \hat x)^\top \Pi^{-1} \Pi^{-1} (x - \mathbfcal{R}_1 \hat x)\\
			& ~~~ + \Big( \frac{\Vert \XdCm Q_1 \Vert^2}{\mu_1} + \frac{\lambda_{\max}(\mathcal{E} \mathcal{E}^\top) \Vert Q_1 \Vert^2}{\mu_2} \Big) \max_{x \in X} \vert \mathcal{N}(x)\vert^2.
		\end{align*}
		Consequently, regarding the fourth term in~\eqref{eq:tmp1}, we can conclude
		\begin{align}
			&2 (x - \mathbfcal{R}_1 \hat x)^\top \Pi^{-1} S  \begin{bmatrix}
					G \\ \bbzero_{n \times (d - n)} \\ \I_{d - n}
				\end{bmatrix}  \mathcal{N}(x) \notag \\
			& \leq (\mu_1 + \mu_2) (x - \mathbfcal{R}_1 \hat x)^\top \Pi^{-1} \Pi^{-1} (x - \mathbfcal{R}_1 \hat x) \notag\\
			& ~~~ + \Big( \frac{\Vert \XdCm Q_1 \Vert^2}{\mu_1} + \frac{\lambda_{\max}(\mathcal{E} \mathcal{E}^\top) \Vert Q_1 \Vert^2}{\mu_2} \Big) \max_{x \in X} \vert \mathcal{N}(x)\vert^2. \phantomsection \label{eq:tmp3}
		\end{align}
		Inspired by the insightful work~\cite{mao2025one}, we now turn our attention to the term ``$\clubsuit$'' in~\eqref{eq:tmp1}. Under the satisfaction of constraint~\eqref{eq: SDP_con3}, one has
		\begin{align}
			\XdCm Q_2 = \XC Q_2 \hat A & \overset{\eqref{eq:data_rel}}{\Longrightarrow} S  \begin{bmatrix}
					\UC \\ \DC
				\end{bmatrix} Q_2 - \EC Q_2 = \XC Q_2 \hat A\notag\\
			& \Longrightarrow S \begin{bmatrix}
					\UC Q_2 \\ \XC Q_2 \\ \pmb{\mathscr{N}} Q_2
				\end{bmatrix} \! - \EC Q_2 = \XC Q_2 \hat A\notag\\
			&  \xRightarrow[\Xi \coloneq \UC Q_2]{\mathbfcal{R}_1 \coloneq \XC Q_2} S \begin{bmatrix}
					\Xi \\ \mathbfcal{R}_1 \\ \pmb{\mathscr{N}} Q_2
				\end{bmatrix} \! - \EC Q_2 = \mathbfcal{R}_1 \hat A  \notag\\
			& \hspace{0.18cm}  \overset{\eqref{eq: SDP_con4}}{\Longrightarrow} S \begin{bmatrix}
					\Xi \\ \mathbfcal{R}_1 \\ \bbzero_{(d - n) \times \hat n}
				\end{bmatrix} \! - \EC Q_2 = \mathbfcal{R}_1 \hat A.  \phantomsection \label{eq:tmp4} 
		\end{align}
		As $\mathbfcal{R}_2 \coloneq \pmb{\mathscr{N}} Q_2 = \bbzero_{(d - n) \times \hat n}$, it is evident that $G \mathbfcal{R}_2 = \bbzero_{m \times \hat n}$, and, accordingly, we can rewrite~\eqref{eq:tmp4} as
		\begin{align}
			 S \begin{bmatrix}
					\Xi - G \mathbfcal{R}_2 \\ \mathbfcal{R}_1 \\ \bbzero_{(d - n) \times \hat n}
				\end{bmatrix} \! - \mathbfcal{R}_1 \hat A =  \EC Q_2, \phantomsection \label{eq:tmp5}
		\end{align}
		showing that the term ``$\clubsuit$" in~\eqref{eq:tmp1} can be replaced by $ \EC Q_2$. Therefore, by using  Cauchy–Schwarz and Young’s inequalities, and since 	\(
		\lambda_{\max}(\EC^\top \EC) \leq \lambda_{\max}(\mathcal{E} \mathcal{E}^\top)
		\), one can deduce the following regarding the second term in~\eqref{eq:tmp1}:
		\begin{align}
			& 2 (x - \mathbfcal{R}_1 \hat x)^\top \Pi^{-1} \Big (S \begin{bmatrix}
					\Xi - G\mathbfcal{R}_2 \\ \mathbfcal{R}_1 \\ \bbzero_{(d - n) \times \hat n}
				\end{bmatrix} - \mathbfcal{R}_1 \hat{A} \Big ) \hat x \notag\\
			& \overset{\eqref{eq:tmp5}}{=} 2 (x - \mathbfcal{R}_1 \hat x)^\top \Pi^{-1} \EC Q_2 \hat x \notag\\
			& \leq \mu_3 (x \! - \! \mathbfcal{R}_1 \hat x)^\top \Pi^{-1} \Pi^{-1} (x \! - \! \mathbfcal{R}_1 \hat x) \! + \! \frac{1}{\mu_3} \hat x^\top Q_2^\top \EC^\top \EC Q_2 \hat x \notag\\
			& \leq \mu_3 (x \! - \! \mathbfcal{R}_1 \hat x)^\top \Pi^{-1} \Pi^{-1} (x \! - \! \mathbfcal{R}_1 \hat x)\notag\\
			& ~~ + \frac{\lambda_{\max}(\mathcal{E} \mathcal{E}^\top) \Vert Q_2 \Vert^2}{\mu_3} \max_{\hat x \in \hat X} \vert \hat x \vert^2.\phantomsection \label{eq:tmp6}
		\end{align}
		We now focus on the term ``$\spadesuit$" in~\eqref{eq:tmp1}. Since $\Psi = \UC Q_3$ and $\mathbfcal{R}_1 \coloneq \XC Q_2$, the term ``$\spadesuit$" under the satisfaction of constraint~\eqref{eq: SDP_con5} can be rewritten as
		\begin{align*}
			& S \begin{bmatrix}
					\Psi \\ \bbzero_{n \times \hat m} \\ \bbzero_{(d - n) \times \hat m}
				\end{bmatrix} - \mathbfcal{R}_1 \hat{B}\\
			& = S \begin{bmatrix}
					\UC Q_3 \\ \DC Q_3
				\end{bmatrix} -\XC Q_2 \hat{B} = S \begin{bmatrix}
					\UC  \\ \DC 
				\end{bmatrix} Q_3 - \XC Q_2 \hat{B}\\
			& \! \overset{\eqref{eq:data_rel}}{=} (\XdCm + \EC) Q_3 - \XC Q_2 \hat B = (\XdCm Q_3 - \XC Q_2 \hat{B}) + \EC Q_3.
		\end{align*}
		Consequently, by using  Cauchy–Schwarz and Young’s inequalities, we can conclude the following regarding the third term in~\eqref{eq:tmp1}:
		\begin{align}
			& 2 (x - \mathbfcal{R}_1 \hat x)^\top \Pi^{-1}  \Big (S \begin{bmatrix}
					\Psi \\ \bbzero_{n \times \hat m} \\ \bbzero_{(d - n) \times \hat m}
				\end{bmatrix} - \mathbfcal{R}_1 \hat{B} \Big) \hat u \notag\\
			& = 2 (x - \mathbfcal{R}_1 \hat x)^\top \Pi^{-1} (\XdCm Q_3 - \XC Q_2 \hat{B}) \hat u \notag\\
			& ~~~ + 2 (x - \mathbfcal{R}_1 \hat x)^\top \Pi^{-1} \EC Q_3 \hat u \notag \\
			& \leq (\mu_4 + \mu_5) (x \! - \! \mathbfcal{R}_1 \hat x)^\top \Pi^{-1} \Pi^{-1} (x \! - \! \mathbfcal{R}_1 \hat x) \notag\\
			& ~~~ + \frac{1}{\mu_4} \Vert \XdCm Q_3 - \XC Q_2 \hat{B} \Vert^2 \vert \hat u \vert^2 + \frac{1}{\mu_5} \hat{u}^\top Q_3^\top \EC^\top \EC Q_3 \hat u \notag\\
			& \leq (\mu_4 + \mu_5) (x \! - \! \mathbfcal{R}_1 \hat x)^\top \Pi^{-1} \Pi^{-1} (x \! - \! \mathbfcal{R}_1 \hat x) \notag\\
			&~~~ + \Big( \frac{\Vert \XdCm Q_3 - \XC Q_2 \hat{B} \Vert^2}{\mu_4} + \frac{\lambda_{\max}(\mathcal{E} \mathcal{E}^\top) \Vert Q_3 \Vert^2}{\mu_5} \Big) \vert \hat u \vert^2. \phantomsection \label{eq:tmp7}
		\end{align}
		Hence, one can rewrite~\eqref{eq:tmp1}, considering~\eqref{eq:tmp3}, \eqref{eq:tmp6}, and~\eqref{eq:tmp7}, as
		\begin{align}
			& \mathscr{L}\CTSF(x,\hat x) \notag\\
			&  \leq (x  -  \mathbfcal{R}_1 \hat x)^\top \Big(  \Pi^{-1} S  \begin{bmatrix}
					K\Pi^{-1} \\ \I_{n} \\ \bbzero_{(d - n) \times n}
				\end{bmatrix}  +  \begin{bmatrix}
					K\Pi^{-1} \\ \I_{n} \\ \bbzero_{(d - n) \times n}
				\end{bmatrix}^{\top} S^\top \Pi^{-1} \notag\\
			&~ +  \sum_{j=1}^{5} \mu_j \Pi^{-1} \Pi^{-1} \Big)  (x \! - \! \mathbfcal{R}_1 \hat x) \! + \! \frac{\lambda_{\max}(\mathcal{E} \mathcal{E}^\top) \Vert Q_2 \Vert^2}{\mu_3} \max_{\hat x \in \hat X} \vert \hat x \vert^2 \notag\\
			&~ +  \Big( \frac{\Vert \XdCm Q_1 \Vert^2}{\mu_1} + \frac{\lambda_{\max}(\mathcal{E} \mathcal{E}^\top) \Vert Q_1 \Vert^2}{\mu_2} \Big) \max_{x \in X} \vert \mathcal{N}(x)\vert^2 \notag\\
			&~ + \Big( \frac{\Vert \XdCm Q_3 - \XC Q_2 \hat{B} \Vert^2}{\mu_4} + \frac{\lambda_{\max}(\mathcal{E} \mathcal{E}^\top) \Vert Q_3 \Vert^2}{\mu_5} \Big) \vert \hat u \vert^2. \phantomsection \label{eq:tmp8}
		\end{align}
		We now solely focus on the first term in~\eqref{eq:tmp8}. In particular, one has
		\begin{align}
			&(x  -  \mathbfcal{R}_1 \hat x)^\top \Big(  \Pi^{-1} S  \begin{bmatrix}
					K\Pi^{-1} \\ \I_{n} \\ \bbzero_{(d - n) \times n}
				\end{bmatrix}  +  \begin{bmatrix}
					K\Pi^{-1} \\ \I_{n} \\ \bbzero_{(d - n) \times n}
				\end{bmatrix}^{\top} S^\top \Pi^{-1} \notag\\
			&\hspace{2.2cm} + \sum_{j=1}^{5} \mu_j \Pi^{-1} \Pi^{-1} \Big)  (x \! - \! \mathbfcal{R}_1 \hat x) \notag\\
			& = (x  -  \mathbfcal{R}_1 \hat x)^\top \Pi^{-1} \Big( S  \begin{bmatrix}
					K \\ \ \Pi \\ \bbzero_{(d - n) \times n}
				\end{bmatrix}  +  \begin{bmatrix}
					K \\ \Pi \\ \bbzero_{(d - n) \times n}
				\end{bmatrix}^{\top} S^\top  \notag\\
			&\hspace{3.2cm} + \sum_{j=1}^{5} \mu_j \I_{n}  \Big) \Pi^{-1} (x  -  \mathbfcal{R}_1 \hat x) \notag\\
			& = (x  -  \mathbfcal{R}_1 \hat x)^\top \Pi^{-1} \begin{bmatrix}
					\I_{n}\\S^\top
				\end{bmatrix}^{\! \top} \begin{bmatrix}
					\sum_{j=1}^{5} \mu_j \I_{n} & \begin{bmatrix}
						K \\ \Pi \\ \bbzero_{(d - n) \times n}
					\end{bmatrix}^{\! \top}\\
					\star & \bbzero_{(m + d)\times (m + d)}
			\end{bmatrix} \notag\\
			& ~~~ \times \begin{bmatrix}
					\I_{n}\\S^\top
				\end{bmatrix} \Pi^{-1} (x  -  \mathbfcal{R}_1 \hat x). \phantomsection \label{eq:tmp9}
		\end{align}
		By defining $\mathrm{s}(x, \hat x) \coloneq \begin{bmatrix}
				\I_{n}\\S^\top
			\end{bmatrix} \Pi^{-1} (x  -  \mathbfcal{R}_1 \hat x)$, we can rewrite~\eqref{eq:tmp9} as
		\begin{align*}
			\mathrm{h}_1(x , \hat x) \coloneq \mathrm{s}^{\top}(x, \hat x) \overbrace{\begin{bmatrix}
						\sum_{j=1}^{5} \mu_j \I_{n} & \begin{bmatrix}
							K \\ \Pi \\ \bbzero_{(d - n) \times n}
						\end{bmatrix}^{\! \top}\\
						\star & \bbzero_{(m + d)\times (m + d)}
				\end{bmatrix}}^{\mathcal{K}_1} \mathrm{s}(x, \hat x).
		\end{align*}
		We also reformulate $\CTSF(x, \hat x)$ as
		\begin{align*}
			\CTSF(x, \hat x) & = (x- \mathbfcal{R}_1\hat x)^\top \Pi^{-1} (x- \mathbfcal{R}_1\hat x)\\
			& = (x  -  \mathbfcal{R}_1 \hat x)^\top \Pi^{-1} \begin{bmatrix}
					\I_{n}\\S^\top
				\end{bmatrix}^{\! \top} \begin{bmatrix}
					\Pi & \bbzero_{n \times (m + d)}\\\star & \bbzero_{(m + d)\times (m + d)}
			\end{bmatrix}  \\
			& ~~~ \times \begin{bmatrix}
					\I_{n}\\S^\top
				\end{bmatrix} \Pi^{-1} (x  -  \mathbfcal{R}_1 \hat x)\\
			& = \mathrm{s}^{\top}(x, \hat x) \overbrace{\begin{bmatrix}
						\Pi & \bbzero_{n \times (m + d)}\\\star & \bbzero_{(m + d)\times (m + d)}
				\end{bmatrix}}^{\mathcal{K}_2}  \mathrm{s}(x, \hat x).
		\end{align*}
		Considering condition~\eqref{eq: con2-def-cont} and~\eqref{eq:tmp8}, if we show that $\mathrm{h}_1(x , \hat x) \leq -\kappa \CTSF(x, \hat x)$, or equivalently
		\begin{align}
			\mathrm{s}^{\top}(x, \hat x) \,  (\mathcal{K}_1 + \kappa \mathcal{K}_2) \,  \mathrm{s}(x, \hat x) \leq 0, \phantomsection \label{eq:tmp10}
		\end{align}
		we can conclude the proof with $\rho$ and $\eta$ as in~\eqref{eq:rho_eta}. However, the key challenges are that $\mathrm{s}(x, \hat x)$ includes the \emph{unknown} system matrices, and, at the same time, enforcing $\mathcal{K}_1 + \kappa \mathcal{K}_2 \preceq 0$ is infeasible since its first diagonal block is positive definite by construction, \emph{i.e.}, $\Pi + \sum_{j=1}^{5} \mu_j \I_{n} \succ 0$. Motivated by this, we revisit Assumption~\ref{assump:noise} and reformulate it as
		\begin{align}
			& \EC \EC^\top \preceq \mathcal{E} \mathcal{E}^\top\notag\\
			& \overset{\eqref{eq:data_rel}}{\Longrightarrow} \big( \XdCm - S \begin{bmatrix}
					\UC \\ \DC
				\end{bmatrix}  \big) \big( \XdCm - S \begin{bmatrix}
					\UC \\ \DC
				\end{bmatrix}\big)^\top \preceq \mathcal{E} \mathcal{E}^\top\notag\\
			& \Longrightarrow \begin{bmatrix}
					\I_{n}\\S^\top
				\end{bmatrix}^\top  \overbrace{\begin{bmatrix}
						\XdCm {\XdCm}^{\! \top} - \mathcal{E} \mathcal{E}^\top & -\XdCm \pmb{\mathds{H}}^\top\\
						\star & \pmb{\mathds{H}}\pmb{\mathds{H}}^\top
				\end{bmatrix}}^{\mathcal{K}_3} \begin{bmatrix}
					\I_{n}\\S^\top
				\end{bmatrix} \preceq 0, \phantomsection \label{eq:tmp11}
		\end{align}
		where $\pmb{\mathds{H}} \coloneq \begin{bmatrix}
				\UC^\top & \DC^\top
			\end{bmatrix}^{\! \top}$. Hence, according to~\eqref{eq:tmp11}, one can straightforwardly conclude that
		\begin{align*}
			\mathrm{h}_2(x, \hat x) \coloneq \mathrm{s}^\top(x, \hat x) \, \mathcal{K}_3 \, \mathrm{s}(x, \hat x) \leq 0.
		\end{align*}
		This is exactly where the classical S-procedure~\cite{polik2007survey} can help us conclude~\eqref{eq:tmp10}. More precisely, according to the classical S-procedure, if there exists a multiplier $\mu_6 \in \Rpz$ such that $\mathrm{h}_1(x , \hat x) +\kappa \CTSF(x, \hat x) - \mu_6 \mathrm{h}_2(x, \hat x) \leq 0$, or equivalently,
		\begin{align}
			\mathrm{s}^\top(x, \hat x) (\mathcal{K}_1 + \kappa \mathcal{K}_2 - \mu_6 \mathcal{K}_3) \, \mathrm{s}(x, \hat x) \leq 0, \label{eq:tmp12}
		\end{align}
		we can conclude that $\mathrm{h}_1(x , \hat x) +\kappa \CTSF(x, \hat x) \leq 0$, and, hence, deduce that~\eqref{eq:tmp10} holds. Evidently, the satisfaction of constraint~\eqref{eq: SDP_con7} implies that $\mathcal{K}_1 + \kappa \mathcal{K}_2 - \mu_6 \mathcal{K}_3 \preceq 0$, meaning that~\eqref{eq:tmp12} holds. Therefore, one can readily conclude that $\mathrm{h}_1(x , \hat x) \leq -\kappa \CTSF(x, \hat x)$. Since $\mathrm{h}_1(x , \hat x)$ is the first term in~\eqref{eq:tmp8}, one has
		\begin{align*}
			\mathscr{L}\CTSF(x,\hat x) \leq -\kappa \CTSF(x, \hat x) + \rho \vert \hat u \vert^2 + \eta,
		\end{align*}
		with $\rho$ and $\eta$ given in~\eqref{eq:rho_eta}, thereby concluding the proof.}
	\end{proof}
	
	\begin{remark}[\textbf{On Choosing $\hat{n}$}]\phantomsection\label{rem:on_n}
		{We note that smaller values of $\hat{n}$ generally simplify the controller design for ct-ROMs.} However, the selection of $\hat{n}$ typically depends on two main factors. First, the specification to be enforced on the original system plays a crucial role. For instance, when the objective is to satisfy a reach-while-avoid specification involving only the first two state variables of a high-dimensional system \BS{(\emph{e.g.}, when the first two states correspond to positions along the x- and y-axes, respectively)}, it is preferable to construct a ct-ROM with two state variables to allow for direct control. Second, it is necessary to determine whether the system dimension can be reduced to the desired value {while maintaining a meaningful closeness guarantee.} If this is not possible, one should progressively consider larger values of $\hat{n}$ until a suitable ct-ROM is obtained. For example, if $\hat{n} = 2$ is desired but unattainable, one may instead consider $\hat{n} = 3$ using the proposed conditions and continue this procedure until a viable choice is identified.
	\end{remark}
	
	\begin{remark}[\textbf{On Matrices $\hat A$ and $\hat{B}$}]\phantomsection \label{Remark_ hatB}
		As is evident, if both $\mathbfcal{R}_1$ and $\hat{A}$ were to be designed simultaneously, constraint~\eqref{eq: SDP_con3} would become bilinear due to their product. This bilinearity is intrinsic to SF-based approaches and arises even in the model-based setting (cf., condition~(20a) in~\cite{zamani2017compositional}). {To address this issue, for a selected value of $\hat{n}$ chosen according to Remark~\ref{rem:on_n},} one can fix the matrix $\hat{A}$ and subsequently solve the SDP~\eqref{eq: SDP}. It is important to note, however, that selecting $\hat{A}$ is not particularly demanding; in fact, to simplify the design process for the ct-ROM $\CTROM$, $\hat{A}$ can be chosen as a Hurwitz matrix, thereby easing the control problem for the ct-ROM $\CTROM$. 
		At the same time, Theorem~\ref{thm: main-cont} imposes no restrictions on the choice of the matrix $\hat{B}$. Consequently, selecting $\hat{B} = \gamma \I_{\hat{n}}$ (\emph{i.e.}, $\hat{m} = \hat{n}$), with $\gamma \in \R \backslash \{0\}$, is a convenient option to ensure that the ct-ROM $\CTROM$ is fully actuated. This choice facilitates the controller synthesis problem, while also enabling regulation of the parameter $\rho$, and consequently the closeness guarantee in~\eqref{eq: error-cont}, through appropriate tuning of $\gamma$, given that $\hat{B}$ directly influences $\rho$ in~\eqref{eq:rho1}. In light of this discussion, choosing a (diagonal) Hurwitz matrix for $\hat{A}$ together with $\hat{B} = \gamma \I_{\hat{n}}$ emerges as an effective design choice, simplifying the controller design for the ct-ROM~$\CTROM$.
	\end{remark}
	
	\subsection{\BS{Extension to Incomplete Dictionaries}}\phantomsection\label{subsec:incomplete_dic}
	\BS{The proposed result in Section~\ref{subsec:dd-ctrom-sfs} holds under the assumption that the dictionary in~\eqref{Dictionary} contains all nonlinear terms present in the dynamics of the ct-NCS in~\eqref{eq: ct-NCS} (cf., Assumption~\ref{assump-on-D}). A natural question, however, concerns the situation in which the dictionary in~\eqref{Dictionary} is incomplete, \emph{i.e.}, when certain nonlinear terms appearing in the true system dynamics are neglected and therefore not included in~\eqref{Dictionary}. In this case, the ct-NCS in~\eqref{eq: ct-NCS} is characterized as
	\begin{align}\phantomsection \label{eq: ct-NCS-incomplete}
		\CT\!: \begin{cases}
				\dot{x} = A_1 x + A_2 \, \mathcal{N}(x) + \mathcal{M}(x) + Bu,\\
				y = x,
		\end{cases}
	\end{align}
	where $\mathcal{M} : X \to \R^{n}$ represents the nonlinear terms that are not considered in the dictionary in~\eqref{Dictionary}. In this section, we develop a data-driven approach to address the presence of such neglected nonlinearities. To this end, we raise the following assumption on $\mathcal{M}(x)$.}
	
	\begin{assumption}[\textbf{Bound on $\mathcal{M}(x)$}]\phantomsection\label{assump:M_bound}
		\BS{A known scalar $\varphi_2 \in \Rp$ satisfies $\vert \mathcal{M}(x) \vert^2 \leq \varphi_2$ for all $x \in X$.}
	\end{assumption}
	
	\begin{remark}[\textbf{On Assumption~\ref{assump:M_bound}}]\phantomsection\label{rem:assumpM}
		\BS{We note that Assumption~\ref{assump:M_bound} is not restrictive, particularly within the scope of this work, where all analyses are carried out over compact state spaces. More specifically, according to Definition~\ref{def: ct-NCS}, since \emph{(i)} $X$ is compact in our analysis and \emph{(ii)} the vector function $f$ is continuously differentiable and therefore continuous, there always exists a scalar $\varphi_2 \in \Rp$ satisfying Assumption~\ref{assump:M_bound}. Nevertheless, it should be noted that potential conservatism may be introduced if the chosen $\varphi_2$ is unnecessarily large, as it directly affects the closeness guarantee~\eqref{eq: error-cont} (cf.,~\eqref{eta2}).}
	\end{remark}
	
	\BS{In light of the ct-NCS in~\eqref{eq: ct-NCS-incomplete}, it can be seen that the term $\dot{x}-\mathbfcal{R}_1\dot{\hat{x}}$ admits a parameterization similar to that in~\eqref{eq:parameterization}; specifically, it is the sum of the terms in~\eqref{eq:parameterization} and the additional term $\mathcal{M}(x)$. Moreover, analogously to~\eqref{eq:data_rel}, one obtains
	\begin{align}
		\XdCm = S  \begin{bmatrix}
				\UC \\ \DC
			\end{bmatrix} - \EC + \MC, \phantomsection\label{eq:data_rel_2}
	\end{align}
	where
	\begin{align*}
		\MC = \begin{bmatrix}
				\mathcal{M}(x(t_0)\!) & 	\!\!\mathcal{M}(x(t_0 + \tau)\!) & \!\!\dots & 	\!\!\mathcal{M}(x(t_0 + (T - 1)\tau)\!)
		\end{bmatrix}
	\end{align*}
	denotes the data matrix associated with the neglected nonlinearities, which is fully \emph{unknown} and is only used for subsequent arguments. Under Assumption~\ref{assump:M_bound}, and following Remark~\ref{rem: Assump-noise}, it follows that the unknown data matrix $\MC$ satisfies
	\begin{align}
		\MC \MC^\top \preceq \mathcal{Z} \mathcal{Z}^\top\!, \phantomsection\label{eq:bound_M_matrix}
	\end{align}
	where $\mathcal{Z} \coloneq \sqrt{\varphi_2 T}\I_{n}$, yielding $\mathcal{Z} \mathcal{Z}^\top = \varphi_2 T \I_{n}$.}
	
	\BS{Having specified the setting in which the dictionary is incomplete, together with the corresponding assumption, we are now in a position to state the following corollary addressing this scenario.}
	
	\begin{corollary}[\textbf{Data-Driven Design with Incomplete Dictionaries}]\phantomsection\label{proposition}
		\BS{Given the ct-NCS $\CT$ in~\eqref{eq: ct-NCS-incomplete} and its ct-ROM $\CTROM$ in~\eqref{eq: ct-ROM}, with $\hat{C} = \mathbfcal{R}_1$, let Assumptions~\ref{assump-on-D}--\ref{assump:M_bound} hold\footnote{\BS{The simultaneous use of Assumptions~\ref{assump-on-D} and~\ref{assump:M_bound} corresponds to the case where the available dictionary captures only part of the nonlinear dynamics, while the remaining terms are absorbed into $\mathcal{M}(x)$.}}. Then, if there exist decision variables $Q_1 \in \R^{T \times (d - n)}$, $Q_2 \in \R^{T \times \hat n}$, $Q_3 \in \R^{T \times \hat m}$, $\Pi \in \R^{n \times n}$, with $\Pi \succ 0$, $K \in \R^{m \times n}$, $\mu_1, \, \mu_2, \, \mu_3, \, \mu_4, \, \mu_5, \, \mu_6 \in \Rp$, and $\mu_7 \in \Rpz$ such that, for some $\vartheta, \, \kappa, \, \bar{\mu}, \, \varepsilon \in \Rp$, $\hat A$, and $\hat B$, the SDP
		\begin{mini!}|s|[2]<b>
			{\substack{\mu_1, \dots, \mu_7,\\ Q_1, Q_2, Q_3,\\ K, \Pi}}{\sum_{i=1}^{3} \! \Vert Q_i \Vert \! - \!\! \sum_{j=1}^{6} \! \mu_j  \! + \! \Vert \XdCm Q_1 \Vert \! + \! \Vert \XdCm Q_3 \! - \! \XC Q_2 \hat B \Vert}
			{\label{eq: SDP2}}{\label{eq: mini_SDP2}}
			\addConstraint{\mu_j \leq \bar{\mu}, \quad \forall j \in \{1, \ldots, 6\},}{ \label{eq: SDP2_con1} }
			\addConstraint{	\text{constraints~\eqref{eq: SDP_con2}--\eqref{eq: SDP_con5}},}{ \label{eq: SDP2_con2} }
			\addConstraint{\begin{bmatrix}
						Z & ~~ \begin{bmatrix}
							K\\
							\Pi\\
							\bbzero_{(d - n) \times n}
						\end{bmatrix}^\top + \mu_7 \XdCm \pmb{\mathds{H}}^\top\\
						\star & -\mu_7 \pmb{\mathds{H}}\pmb{\mathds{H}}^\top
					\end{bmatrix} \! \! \preceq 0}{ \label{eq: SDP2_con7} }
		\end{mini!}
		has a solution, where $\pmb{\mathds{H}} \coloneq \begin{bmatrix}
				\UC^\top & \DC^\top
			\end{bmatrix}^{\! \top}$,
		\begin{align*}
			Z \coloneq \kappa \Pi + \sum_{j=1}^{6}  \mu_j  \I_n - \mu_7 (\XdCm {\XdCm}^{\! \top} - \mathcal{Y} ),
		\end{align*}
		and $\mathcal{Y} \coloneq (1 + \vartheta) \mathcal{E}\mathcal{E}^\top + (1 + \frac{1}{\vartheta}) \mathcal{Z} \mathcal{Z}^\top$,
		then $\CTSF(x,\hat{x}) = (x - \mathbfcal{R}_1 \hat{x})^\top P (x - \mathbfcal{R}_1 \hat{x})$ is an SF from $\CTROM$  to $\CT$, with $\mathbfcal{R}_1 \coloneq \XC Q_2$, $P \coloneq \Pi^{-1}$, $\alpha \coloneq \lambda_{\min}(P)$, and
		\begin{subequations}\phantomsection \label{eq:rho_eta2}
			\begin{align}
				\rho &  \coloneq  \frac{\Vert \XdCm Q_3 - \XC Q_2 \hat{B} \Vert^2}{\mu_4} + \frac{\Upsilon \Vert Q_3 \Vert^2}{\mu_5},\phantomsection\label{rho2}\\
				\eta &  \coloneq  \Big( \frac{\Vert \XdCm Q_1 \Vert^2}{\mu_1} + \frac{\Upsilon \Vert Q_1 \Vert^2}{\mu_2} \Big) \max_{x \in X} \vert \mathcal{N}(x) \vert^2 + \frac{\Upsilon \Vert Q_2 \Vert^2}{\mu_3}\notag\\
				& \hspace{0.4cm} \times \max_{\hat x \in \hat X} \vert \hat x \vert^2 + \frac{\varphi_2}{\mu_6} ,\phantomsection\label{eta2}
			\end{align}
			where $\Upsilon \coloneq  \big(\sqrt{\lambda_{\max}(\mathcal{E} \mathcal{E}^\top)} + \sqrt{\lambda_{\max}(\mathcal{Z} \mathcal{Z}^\top)}\big)^2 $.
		\end{subequations}
		Moreover, the interface function is designed in the form of~\eqref{eq:interface}, with $G \coloneq \UC Q_1$, $\mathbfcal{R}_2 \coloneq \pmb{\mathscr{N}} Q_2 = \bbzero_{(d - n) \times \hat n}$, $\Xi \coloneq \UC Q_2$, and $\Psi \coloneq \UC Q_3$.}
	\end{corollary}
	
	\begin{proof}
		\BS{The proof proceeds in a manner analogous to that of Theorem~\ref{thm: main-cont}; therefore, only a sketch of the proof is provided due to space limitations. In particular, the first step is identical and hence omitted here, whereas the second step requires further developments. By taking the Lie derivative of $\CTSF$ along the dynamics in~\eqref{eq: ct-NCS-incomplete}, the terms in~\eqref{eq:tmp1} yield an additional contribution given by $2 (x - \mathbfcal{R}_1 \hat x)^\top \Pi^{-1} \mathcal{M}(x)$. Now, under the satisfaction of constraint~\eqref{eq: SDP_con6}, and in view of~\eqref{eq:data_rel_2}, the term ``$\bigstar$'' in~\eqref{eq:tmp1} can be expressed as $(\XdCm + \EC - \MC) Q_1$ $ \mathcal{N}(x)$. Consequently, regarding the fourth term in~\eqref{eq:tmp1}, one has
			\begin{align*}
				& 2 (x - \mathbfcal{R}_1 \hat x)^\top \Pi^{-1} (\XdCm + \EC - \MC) Q_1 \mathcal{N}(x) \\
				& =   2 (x   -   \mathbfcal{R}_1 \hat x)^{    \top} \Pi^{-1} \XdCm Q_1 \mathcal{N}(x)\\
				& ~~~ + 2 (x   -   \mathbfcal{R}_1 \hat x)^{    \top} \Pi^{-1} (\EC - \MC) Q_1 \mathcal{N}(x)\\
				& \leq  (\mu_1 + \mu_2) (x - \mathbfcal{R}_1 \hat x)^\top \Pi^{-1} \Pi^{-1} (x - \mathbfcal{R}_1 \hat x)\\
				& ~~~ + \frac{1}{\mu_1} \Vert \XdCm Q_1 \Vert^2 \vert \mathcal{N}(x) \vert^2 + \frac{1}{\mu_2} \Vert (\EC - \MC)Q_1 \Vert^2 \vert \mathcal{N}(x) \vert^2.
			\end{align*}
			Since
			\begin{align*}
				\Vert (\EC - \MC)Q_1 \Vert^2  & =  \Vert \EC Q_1 - \MC Q_1 \Vert^2 \\ & \leq  (\Vert \EC Q_1 \Vert + \Vert \MC Q_1 \Vert)^2,
			\end{align*}
			and given that $\Vert \EC Q_1 \Vert \leq \sqrt{\lambda_{\max}(\mathcal{E} \mathcal{E}^\top)} \Vert Q_1 \Vert$ and $\Vert \MC Q_1 \Vert \leq \sqrt{\lambda_{\max}(\mathcal{Z} \mathcal{Z}^\top)} \Vert Q_1 \Vert$,
			one has
			\begin{align*}
				\Vert (\EC - \MC)Q_1 \Vert^2 \! \leq \!  \big( \sqrt{\lambda_{\max}(\mathcal{E} \mathcal{E}^\top)} + \sqrt{\lambda_{\max}(\mathcal{Z} \mathcal{Z}^\top)} \big)^2 \Vert Q_1 \Vert^2  .
			\end{align*}
			Hence, the fourth term in~\eqref{eq:tmp1} can be bounded as
			\begin{align*}
				& 2 (x - \mathbfcal{R}_1 \hat x)^\top \Pi^{-1} (\XdCm + \EC - \MC) Q_1 \mathcal{N}(x)\\
				& \leq (\mu_1 + \mu_2) (x - \mathbfcal{R}_1 \hat x)^\top \Pi^{-1} \Pi^{-1} (x - \mathbfcal{R}_1 \hat x)\\
				& ~~~ + \Big( \frac{\Vert \XdCm Q_1 \Vert^2}{\mu_1} + \frac{\Upsilon \Vert Q_1 \Vert^2}{\mu_2} \Big) \max_{x \in X} \vert \mathcal{N}(x) \vert^2,
			\end{align*}
			where $\Upsilon \coloneq  \big(\sqrt{\lambda_{\max}(\mathcal{E} \mathcal{E}^\top)} + \sqrt{\lambda_{\max}(\mathcal{Z} \mathcal{Z}^\top)}\big)^2 $. One may proceed analogously to derive bounds for the terms ``$\clubsuit$'' and ``$\spadesuit$'' in~\eqref{eq:tmp1}, which are omitted for brevity. Concerning the additional term $2 (x - \mathbfcal{R}_1 \hat x)^\top \Pi^{-1} \mathcal{M}(x)$, by invoking the bound $\vert \mathcal{M}(x) \vert^2 \leq \varphi_2$ from Assumption~\ref{assump:M_bound}, one directly obtains
			\begin{align*}
				& 2 (x - \mathbfcal{R}_1 \hat x)^\top \Pi^{-1} \mathcal{M}(x)\\
				& \leq \mu_6 (x - \mathbfcal{R}_1 \hat x)^\top \Pi^{-1} \Pi^{-1} (x - \mathbfcal{R}_1 \hat x) + \frac{\varphi_2}{\mu_6}.
			\end{align*}
			Following these steps with a similar reasoning, we obtain a bound analogous to that in~\eqref{eq:tmp8}, with $\rho$ and $\eta$ defined as in~\eqref{eq:rho_eta2}. The remainder of the proof proceeds analogously to that of Theorem~\ref{thm: main-cont}, except for the part concerning~\eqref{eq:tmp11}. Specifically, in the S-procedure step, both Assumptions~\ref{assump:noise} and~\ref{assump:M_bound} must now be considered. To this end, for any $\vartheta \in \Rp$, we have
			\[
			(\EC - \MC) (\EC - \MC)^\top \preceq (1 + \vartheta) \EC\EC^\top + (1 + \frac{1}{\vartheta}) \MC \MC^\top,
			\]
			and by invoking~\eqref{eq: noise_bound_matrix} and~\eqref{eq:bound_M_matrix}, it follows that
			\begin{align*}
				& (\EC - \MC) (\EC - \MC)^\top \preceq \overbrace{(1 + \vartheta) \mathcal{E}\mathcal{E}^\top + (1 + \frac{1}{\vartheta}) \mathcal{Z} \mathcal{Z}^\top}^{\mathcal{Y}}\\
				& \overset{\eqref{eq:data_rel_2}}{\Longrightarrow} \big( \XdCm - S \begin{bmatrix}
					\UC \\ \DC
				\end{bmatrix}  \big) \big( \XdCm - S \begin{bmatrix}
					\UC \\ \DC
				\end{bmatrix}\big)^\top \preceq \mathcal{Y}\\
				& \Longrightarrow \! \! \begin{bmatrix}
					\I_{n}\\S^\top
				\end{bmatrix}^{\! \top} \!\!  \begin{bmatrix}
					\XdCm {\XdCm}^{\! \top} - \mathcal{Y} & -\XdCm \pmb{\mathds{H}}^\top\\
					\star & \pmb{\mathds{H}}\pmb{\mathds{H}}^\top
				\end{bmatrix} \!\! \begin{bmatrix}
					\I_{n}\\S^\top
				\end{bmatrix} \preceq 0,
			\end{align*}
			where $\pmb{\mathds{H}} \coloneq \begin{bmatrix}
				\UC^\top & \DC^\top
			\end{bmatrix}^{\! \top}$. The remainder of the proof follows that of Theorem~\ref{thm: main-cont} and is omitted. }
	\end{proof}
	
	\begin{remark}[\textbf{Computing $\eta$}]\phantomsection\label{rem:eta_comp}
		\BS{From~\eqref{eq:eta1} and~\eqref{eta2}, computing $\eta$ requires an upper bound on $\max_{x \in X} \vert \mathcal{N}(x) \vert^2$. Since $X$ is compact and $\mathcal{N}$ is continuous, this maximum is well defined and attained. In simple cases, it can be computed analytically by bounding each component over $X$, \emph{i.e.}, $\max_{x \in X} \vert \mathcal{N}(x) \vert^2 \leq \sum_{i = 1}^{d-n} \max_{x \in X} \mathcal{N}_i^2(x)$, where $\mathcal{N}_i(x)$ denotes the $i$-th component of $\mathcal{N}(x)$. Alternatively, one may compute/approximate $\max_{x \in X} \vert \mathcal{N}(x) \vert^2$ numerically using \texttt{fmincon} in \textsc{Matlab} with multiple initializations and taking the largest value found.
        We emphasize that the dependence of $\eta$ in Theorem~\ref{thm: main-cont} and Corollary~\ref{proposition}, and consequently of the closeness guarantee, on $X$ stems from the generality of our framework, which accommodates arbitrary continuously differentiable nonlinearities in $\mathcal{N}(x)$; their effect is captured through $\eta$ (via $\max_{x \in X} |\mathcal{N}(x)|^2$) and is therefore reflected in the resulting closeness guarantee. If the nonlinearities are inherently bounded, \emph{e.g.}, sinusoidal nonlinearities, then $\max_{x \in X} |\mathcal{N}(x)|^2$ is independent of the choice and size of the state space $X$, implying that the resulting closeness guarantee is unaffected by $X$ and depends solely on the intrinsic bound of the nonlinearities. Moreover, if the underlying dynamics are linear, \emph{i.e.}, $\mathcal{N}(x) = \bbzero_{d-n}$, then $\eta$ becomes significantly smaller without any dependence on $\mathcal{N}(x)$; consequently, a substantially tighter closeness guarantee can be achieved.}
	\end{remark}
	
		\begin{algorithm}[t!]
		\caption{Data-driven construction of SF and interface function}\phantomsection\label{Alg:1}
		\begin{center}
			\begin{algorithmic}[1]
				\REQUIRE 
				\BS{Assumptions~\ref{assump-on-D}--\ref{assump:M_bound}, $\varphi_1$, and $\varphi_2$}
				\STATE 
				\BS{Collect $\UC$, $\XC$, and form $\XdCm$, $\pmb{\mathscr{N}}$, and $\DC$ according to Section~\ref{subsec:data_collection}}
				\STATE
				\BS{Set $\hat n$ according to Remark~\ref{rem:on_n}}
				\STATE
				\BS{Set $\hat A$ and $\hat{B}=\gamma \I_{\hat{n}}$, with $\gamma\in\R\backslash\{0\}$, according to Remark \ref{Remark_ hatB}}\phantomsection\label{Step3}
				\IF{\BS{$\mathcal{D}(x)$ is complete}}
				\STATE
				\BS{For a fixed $\kappa$, $\bar{\mu}$, and $\varepsilon$, solve the SDP~\eqref{eq: SDP}}
				\STATE
				\BS{Compute $\mu_1$--$\mu_6$, $\Pi$, $K$, $Q_1$, $Q_2$, and $Q_3$}
				\STATE
				\BS{Obtain $\mathbfcal{R}$, $P$, $G$, $\Xi$, $\Psi$, $\rho$, and $\eta$ according to Theorem~\ref{thm: main-cont}}
				\ELSE 
				\STATE
				\BS{For a fixed $\vartheta$, $\kappa$, $\bar{\mu}$, and $\varepsilon$, solve the SDP~\eqref{eq: SDP2}}
				\STATE
				\BS{Compute $\mu_1$--$\mu_7$, $\Pi$, $K$, $Q_1$, $Q_2$, and $Q_3$}
				\STATE
				\BS{Obtain $\mathbfcal{R}$, $P$, $G$, $\Xi$, $\Psi$, $\rho$, and $\eta$ according to Corollary~\ref{proposition}}
				\ENDIF
				\STATE
				\BS{Setting $\hat C = \mathbfcal{R}_1$, quantify the closeness between the output trajectories of ct-NCS $\CT$ and ct-ROM $\CTROM$ according to~\eqref{eq: error-cont}}
				\ENSURE
				\BS{SF $\CTSF(x, \hat x) = (x- \mathbfcal{R}_1\hat x)^\top P (x- \mathbfcal{R}_1\hat x)$ and the interface function $u = [KP ~~~ G] (\mathcal{D}(x) - \mathbfcal{R} \hat x) + \Xi \hat x + \Psi \hat u$}
			\end{algorithmic}
		\end{center}
	\end{algorithm}
	
	\begin{remark}[\textbf{On Sampling Time}]\phantomsection\label{rem:samp}
		\BS{The proposed framework, in line with many works in the literature (\emph{e.g.},~\cite{guo2021data}), does not impose any restriction on the sampling time. Specifically, the sampling interval $\tau$ is arbitrary, and the results remain valid even with unevenly spaced measurements. Nevertheless, two related aspects merit attention. First, the noise in the state-derivative data originates from derivative approximation, which introduces an error proportional to $\tau$ (and is therefore treated as noise in the collected data); larger values of $\tau$ may thus lead to higher-amplitude noise. Second, this increase in noise enlarges the entries of $\mathcal{E}\mathcal{E}^\top$ in Assumption~\ref{assump:noise}, which may affect the feasibility of constraints~\eqref{eq: SDP_con7} and~\eqref{eq: SDP2_con7} through $Z$. Consequently, although the results hold irrespective of $\tau$, smaller sampling intervals can improve feasibility, provided Assumption~\ref{assumpt:full-rank} holds. From a practical standpoint, systems with fast dynamics typically require smaller $\tau$~\cite[Section~IV-D]{10565947}.}
	\end{remark}
	
	\BS{It is worth emphasizing that a fundamental and shared feature of Theorem~\ref{thm: main-cont} and Corollary~\ref{proposition} is that the system’s unknown matrices are never identified. To be more precise, the matrix
		\(S =
		\begin{bmatrix}
			B & \! \! A_1 & \!\! A_2
		\end{bmatrix}\)
		is neither estimated nor does it appear in the proposed conditions. Instead, by leveraging the classical S-procedure, the proposed design establishes correctness guarantees for all systems that are consistent with the measured data and the assumed noise bounds.} Algorithm~\ref{Alg:1} summarizes the steps of the proposed data-driven framework.
	
	\section{\BS{Discussion}}\phantomsection\label{Sec:Discussion}
	\BS{This section discusses several key aspects of the proposed data-driven framework, including insights into the proposed conditions and their feasibility, a special case yielding the smallest closeness error, an extension to systems subject to process disturbances, an analysis of computational complexity, and a discussion of underlying limitations.}
	
	\subsection{\BS{On the Proposed Conditions}}\phantomsection\label{Subsec:Disc1}
	\BS{Here, we focus on Theorem~\ref{thm: main-cont} as the primary result, noting that Corollary~\ref{proposition} builds upon this theorem and can be interpreted through similar reasoning. Let us begin by discussing the cost function in~\eqref{eq: mini_SDP}. We recall that smaller values of $\rho$ and $\eta$ yield a tighter closeness guarantee in~\eqref{eq: error-cont}.
	With this in mind, and in view of the expressions of $\rho$ and $\eta$ in~\eqref{eq:rho_eta}, it can be observed that minimizing $\sum_{i=1}^{3} \! \Vert Q_i \Vert$ contributes to reducing both $\rho$ and $\eta$, as the optimization biases the solution toward matrices $Q_i$ with smaller norms, thereby yielding a tighter closeness guarantee.
	Similarly, minimizing $\Vert \XdCm Q_3 \! - \! \XC Q_2 \hat B \Vert$ and $\Vert \XdCm Q_1 \Vert$ contributes to reducing $\rho$ and $\eta$, respectively, thereby further tightening the closeness guarantee.
	The term $-\sum_{j=1}^{5} \mu_j$ is included to bias the solution toward larger values of $\mu_1$–$\mu_5$, which in turn leads to smaller values of $\rho$ and $\eta$, and consequently to a tighter closeness guarantee.
	In connection with this maximization (\emph{i.e.,} minimizing $-\sum_{j=1}^{5} \mu_j$), note that constraint~\eqref{eq: SDP_con1} is imposed solely to prevent the optimization problem from becoming unbounded; accordingly, $\bar{\mu}$ can be selected sufficiently large. Collectively, it is evident that the cost function in~\eqref{eq: mini_SDP} significantly helps tighten the closeness guarantee~\eqref{eq: error-cont}. We note that while each application of Young’s inequality introduces parameters $\mu_j$, which can potentially add conservatism to the resulting closeness guarantee, they are maximized within the optimization problem, thereby mitigating this conservatism.}
	
	\BS{While the cost function in~\eqref{eq: mini_SDP} directly aims at minimizing $\rho$ and $\eta$ to obtain tight closeness guarantees, another parameter that plays a crucial role in the resulting guarantee is $\alpha = \lambda_{\min}(P)$. Hence, the cost function in~\eqref{eq: mini_SDP} can be further refined to incorporate the maximization of $\alpha$, thereby yielding a tighter closeness guarantee. To this end, one can introduce a new decision variable $\zeta \in \Rp$ and add the constraint $\Pi \preceq \zeta \I_n$ to the SDP~\eqref{eq: SDP}, while minimizing $\zeta$ through its inclusion in the cost function. This procedure can potentially lead to a smaller $\lambda_{\max}(\Pi)$. Since $P = \Pi^{-1}$, it follows that $\lambda_{\min}(P) = \frac{1}{\lambda_{\max}(\Pi)}$. Therefore, reducing $\lambda_{\max}(\Pi)$ results in a larger $\alpha$ (as $\alpha = \lambda_{\min}(P)$) and potentially tighter closeness guarantees.}
	
	\BS{It is worth noting that, although the SDP~\eqref{eq: SDP} can be solved as a single optimization problem, the use of a summed objective does not necessarily enforce the strongest possible minimization of each individual term in the cost function. In fact, the optimizer minimizes the aggregate objective and may therefore trade off improvements across different terms, rather than pushing each component toward its individual optimum. Owing to the structural separability of the decision variables and the associated constraints, the problem can be reformulated as three independent subproblems. In particular, the constraints involving $Q_1$, those involving $(Q_2, Q_3)$, and those depending on $(K, \Pi, \mu_j)$ are mutually disjoint, while the objective function is additively separable with respect to these variable groups. Consequently, the original SDP can be decomposed into three smaller SDPs, each comprising the corresponding subset of decision variables together with its associated cost functions and constraints. This decomposition enables a more targeted minimization of the individual objective components and can further improve numerical conditioning, as well as the interpretability of the resulting design parameters.}
	
	\BS{We now discuss the constraints in the SDP~\eqref{eq: SDP}. Concerning~\eqref{eq: SDP_con6}, it is feasible provided that $\operatorname{rank}(\DC) = d$, which constitutes a \emph{sufficient} condition for feasibility. More precisely, if $\operatorname{rank}(\DC) = d$, then every vector in $\R^d$, and hence every column of
		\(
		\begin{bmatrix}
			\bbzero_{n \times (d - n)} \\ \I_{d - n}
		\end{bmatrix}
		\),
		lies in $\operatorname{col}(\DC)$, thereby rendering constraint~\eqref{eq: SDP_con6} feasible. This shows that, under Assumption~\ref{assumpt:full-rank}, constraint~\eqref{eq: SDP_con6} is feasible.
		As a simple example, one can observe that
		\[
		Q_1 \coloneq \DC^\top (\DC \DC^\top)^{-1} \begin{bmatrix}
			\bbzero_{n \times (d - n)} \\ \I_{d - n}
		\end{bmatrix}
		\]
		is always a valid solution, since $\DC \DC^\top$ is invertible under the assumption that $\operatorname{rank}(\DC) = d$. Nevertheless, this choice is not necessarily the most suitable one; in particular, the desired solution should also minimize $\Vert \XdCm Q_1 \Vert$, which effectively attenuates the influence of nonlinearities (cf.,~\eqref{eq:tmp13}) and thereby tightens the closeness guarantee, as discussed previously.}
		
		\BS{We also note that, since $Q_1 \in \R^{T \times (d-n)}$ must be determined so as to minimize $\Vert Q_1 \Vert$ and $\Vert \XdCm Q_1 \Vert$ while satisfying constraint~\eqref{eq: SDP_con6}, increasing the amount of collected data can facilitate this design task. In fact, a larger number of samples increases the number of decision variables, thereby providing additional degrees of freedom that can improve the attainable optimization performance. However, collecting excessive data introduces additional computational burden and may adversely affect feasibility, since the entries of the matrix $\mathcal{E} \mathcal{E}^\top$ in Assumption~\ref{assump:noise} may become larger as the number of samples increases. This highlights the need to balance data richness, computational complexity, and feasibility considerations throughout the design process.}
	
	\BS{As is evident, $Q_2 = \bbzero_{T \times \hat n}$ is a trivial but undesirable solution to constraints~\eqref{eq: SDP_con3} and~\eqref{eq: SDP_con4}. For this reason, constraint~\eqref{eq: SDP_con2}, with $\varepsilon \in \Rp$, which can be arbitrarily small, is imposed to preclude the trivial solution. We emphasize that this constraint is not unique, and alternative constraints may be employed, provided that feasibility is preserved. For instance, one may impose constraints directly on $\mathbfcal{R}_1$ by enforcing suitable conditions on $\XC Q_2$, such as fixing a selected entry of $\XC Q_2$ to be equal to one, thereby excluding the trivial solution. In contrast, although $Q_3 = \bbzero_{T \times \hat m}$ is a trivial solution to constraint~\eqref{eq: SDP_con5}, no additional constraint is required to prevent its selection. This is because, by minimizing $\Vert \XdCm Q_3 \! - \! \XC Q_2 \hat B \Vert$, the solver is naturally encouraged to select a nontrivial solution (recalling that $Q_2 \neq \bbzero_{T \times \hat n}$) while still satisfying constraint~\eqref{eq: SDP_con5}. Notice that, under Assumption~\ref{assumpt:full-rank}, constraints~\eqref{eq: SDP_con4} and~\eqref{eq: SDP_con5} are feasible, following the preceding discussion on the feasibility of constraint~\eqref{eq: SDP_con6}.}
	
	\BS{We also note that, for a fixed matrix $\hat{A}$, constraint~\eqref{eq: SDP_con3} alone is a homogeneous system of linear equations and is therefore always feasible, since it admits the trivial solution $Q_2 = \bbzero_{T \times \hat n}$. Moreover, constraint~\eqref{eq: SDP_con3} contains $n\hat{n}$ scalar equalities and $T\hat{n}$ scalar decision variables. Hence, the dimension of the solution space is at least $T\hat{n} - n\hat{n} = (T - n)\hat{n}$. In particular, if $T > n$, then $(T - n)\hat{n} > 0$, and constraint~\eqref{eq: SDP_con3} admits nontrivial solutions.  
	As Assumption~\ref{assumpt:full-rank} implies $T \geq m + d$ (cf., Remark~\ref{rem:full-data}), and $m + d > n$ (when $m \geq 1$), one can conclude that constraint~\eqref{eq: SDP_con3} is feasible under Assumption~\ref{assumpt:full-rank}, while admitting a nontrivial solution. We also remark that Assumption~\ref{assumpt:full-rank} prevents a structural obstruction potentially caused by a singular bottom-right block in constraint~\eqref{eq: SDP_con7}, as it ensures that $\pmb{\mathds{H}}\pmb{\mathds{H}}^\top \succ 0$. Indeed, constraint~\eqref{eq: SDP_con7} may still be satisfied even if Assumption~\ref{assumpt:full-rank} does not hold; however, this would require each column of the block denoted by $\star$ to lie within $\operatorname{col}(\pmb{\mathds{H}}\pmb{\mathds{H}}^\top)$, which is nontrivial and constitutes a strong additional restriction. When Assumption~\ref{assumpt:full-rank} holds, this requirement is bypassed, thereby facilitating the satisfaction of constraint~\eqref{eq: SDP_con7}.}
	
	\subsection{\BS{Special Case: Least Closeness Error}}\phantomsection\label{subsec:SC}
	\BS{The proposed data-driven framework can offer additional benefits in a special setting. This corresponds to the case in which the state-derivative data in~\eqref{eq: XdC} are assumed to be accurately available, \emph{i.e.}, not corrupted by noise, the dictionary is complete, and the nonlinearities are matched (\emph{i.e.}, $A_2 \, \mathcal{N}(x) \in \operatorname{col}(B)$ for all $x \in X$). This implies the existence of a vector-valued map $\aleph : X \to U$ such that $A_2 \, \mathcal{N}(x) = B \aleph(x)$, meaning that all system nonlinearities can be directly canceled when the dictionary is complete, \emph{i.e.}, the solution can satisfy $\Vert \XdCm Q_1 \Vert = 0$. 
	Accordingly, since there is no source of noise in the data and the dictionary is complete, $\eta$ in~\eqref{eq:eta1} can be obtained as zero, while $\rho$ in~\eqref{eq:rho1} only consists of its first term. Collectively, these properties lead to the least conservative bound within the present framework. However, while certain practical systems possess the matched-nonlinearities property (\emph{e.g.}, aerial vehicles, flexible-joint manipulators, cable-driven manipulators, and mobile agents, as reported in~\cite{7927430,su2011adaptive}), this assumption can be generally restrictive in practice.
	This constitutes the main motivation for minimizing $\Vert \XdCm Q_1 \Vert$ rather than enforcing $\XdCm Q_1 = \bbzero_{n \times (d-n)}$, thereby accommodating unmatched nonlinearities and covering a broader class of applications, albeit at the expense of looser closeness guarantees.}
	
	\subsection{\BS{Extension to Systems with Process Disturbances}}\phantomsection\label{subsec:Disturbance}
	\BS{The proposed framework can be readily extended to systems subject to process disturbances. More specifically, by imposing an assumption on the disturbance magnitude, analogous to Assumption~\ref{assump:M_bound}, and following an approach similar to that in Section~\ref{subsec:incomplete_dic}, the framework can be adapted accordingly to simultaneously account for robustness with respect to incomplete dictionaries, state-derivative noise, and process disturbances. While promising, the immediate consequences of such an extension are that $\eta$ in~\eqref{eta2} becomes explicitly dependent on the disturbance bound, $\Upsilon$ is modified accordingly, and $Z$ in constraint~\eqref{eq: SDP2_con7} is likewise affected by the disturbance bound, now comprising three terms corresponding to the three sources of noise/uncertainty in the collected data. This illustrates that extending the approach to systems subject to process disturbances is conceptually straightforward, albeit at the expense of increased conservatism and a looser closeness bound. It is worth highlighting that if only process disturbances and noise in the state-derivative data are considered, then, instead of $\mathcal{M}(x)$, the process disturbance would appear explicitly in the dynamics, and therefore the result in Section~\ref{subsec:incomplete_dic} can be directly applied.}
	
	\subsection{\BS{Computational Complexity Analysis}}\phantomsection\label{subsec:CC}
	\BS{Here, we provide a computational complexity analysis for Theorem~\ref{thm: main-cont}, noting that a similar reasoning extends to Corollary~\ref{proposition}. Inspecting the SDP~\eqref{eq: SDP}, one can readily observe that constraint~\eqref{eq: SDP_con7} is an $\mathrm{L}_1 \times \mathrm{L}_1$ linear matrix inequality (LMI), where $\mathrm{L}_1 \coloneq n + m + d$. 
	Furthermore, each term in the cost function~\eqref{eq: mini_SDP}, aiming to minimize the induced $2$-norm of a matrix, introduces an additional LMI of an appropriate dimension. As a result, there are five additional LMIs with dimensions $\mathrm{L}_i \times \mathrm{L}_i$ for all $i \in \{2,\ldots,6\}$. Specifically, $\mathrm{L}_2 = T + d - n$, $\mathrm{L}_3 = T + \hat n$, $\mathrm{L}_4 = T + \hat m$, $\mathrm{L}_5 = d$, and $\mathrm{L}_6 = n + \hat m$. For convenience, we define $\mathrm{N} \coloneq \sum_{i=1}^{6} \mathrm{L}_i$.
	Moreover, the SDP~\eqref{eq: SDP} involves
	\[
	\mathrm{M} \coloneq \overbrace{d(d-n)}^{\eqref{eq: SDP_con6}} + \overbrace{n\hat n}^{\eqref{eq: SDP_con3}} + \overbrace{(d-n)\hat n}^{\eqref{eq: SDP_con4}} + \overbrace{d\hat m}^{\eqref{eq: SDP_con5}}
	\]
	scalar equality constraints; constraints~\eqref{eq: SDP_con1} and~\eqref{eq: SDP_con2} are excluded from this count since their sizes remain unchanged, and they do not affect asymptotic scaling. Finally, the number of scalar decision variables is given by
	\[
	\mathrm{V} \coloneq \overbrace{\frac{n(n+1)}{2}}^{\Pi} + \overbrace{mn}^{K} + \overbrace{T(d-n)}^{Q_1} + \overbrace{T\hat n}^{Q_2} + \overbrace{T\hat m}^{Q_3},
	\]
	excluding $\mu_1$ to $\mu_6$ as their number does not affect asymptotic scaling.}
	
	\BS{At the same time, a rough time complexity estimate for dense primal-dual interior-point methods is
	\begin{align}
		\text{Time} = \mathcal{O}\Big(\sqrt{\mathrm{N}}\,\log\!\frac{1}{\epsilon}\;\big(\mathrm{M}^3 + \mathrm{M}^2\mathrm{N}^2 + \mathrm{M} \mathrm{N}^3 + \mathrm{V}^3\big)\Big), \phantomsection\label{eq:time}
	\end{align}
	where $\epsilon$ denotes the solution accuracy (typically on the order of $10^{-8}$). Moreover, a rough memory complexity estimate is
	\begin{align}
		\text{Memory} = \mathcal{O}\big(\mathrm{M}^2 + \mathrm{N}^2 + \mathrm{V}^2 + \mathrm{M}\mathrm{N}^2\big). \phantomsection\label{eq:mem}
	\end{align}
	In the common regime where $T > d \gg n \gg \hat n$ and $n \geq m \geq \hat m$, one has $\mathrm{N} \asymp T$, $\mathrm{M} \asymp d^2$, and $\mathrm{V} \asymp Td$, where
	$\mathrm{N} \asymp T$ means that $\mathrm{N}$ and $T$ grow at the same rate, up to constant factors.
	Consequently, for~\eqref{eq:time}, considering $T \ngeq d^2$ as the primary subregime, we obtain $\mathrm{M}^3 + \mathrm{M}^2\mathrm{N}^2 + \mathrm{M} \mathrm{N}^3 + \mathrm{V}^3 \asymp T^3 d^3 + d^4 T^2$ and $\sqrt{\mathrm{N}} \asymp T^{0.5}$, while for~\eqref{eq:mem}, $\mathrm{M}^2 + \mathrm{N}^2 + \mathrm{V}^2 + \mathrm{M}\mathrm{N}^2 \asymp d^2 T^2$. 
	Note that, according to Remark~\ref{rem:full-data}, although $T \geq d + m$ is required, selecting $T \geq d^2$ may be unnecessarily large; this justifies the assumption $T \not\geq d^2$ in order to facilitate a simpler analysis. To further simplify the exposition, let us assume $T \asymp d$. Then, we obtain
	\[
	\text{Time} \approx \mathcal{O}\big(d^{6.5} \log\!\tfrac{1}{\epsilon}\big), 
	\qquad 
	\text{Memory} \approx \mathcal{O}\big(d^4\big),
	\]
	which is comparable to the bounds reported in~\cite{8619019}. Therefore, following~\cite{8619019}, one can conclude that the proposed framework can be employed for systems with dictionaries containing no more than a few hundred elements. This complexity analysis highlights an important aspect of the proposed methodology. Specifically, the nonlinear vector function $f(x)$ is lifted into a higher-dimensional feature space defined by the dictionary $\mathcal{D}(x)$ in~\eqref{Dictionary}, yielding a linear-in-features parameterization of $f(x)$. As is clear from Theorem~\ref{thm: main-cont} and Corollary~\ref{proposition}, the proposed conditions depend on the length of this dictionary, \emph{i.e.}, $d$, which may be considerably larger than the state dimension $n$. The preceding analysis therefore indicates that accommodating large dictionaries comes at the price of increased computational complexity, as expected.}
	
	\subsection{\BS{Underlying Limitations}}\phantomsection\label{subsec:lim}
	\BS{Similar to any methodology, the proposed framework has certain limitations, which may also be viewed as directions for future research. First, the framework assumes the direct measurability of all state variables, which is required to collect input--state data~\eqref{eq: data_ct} from the system. This assumption can be restrictive in practical scenarios where only partial output measurements are available. This motivates extending the framework to handle such cases, which is a challenging task. Indeed, even when the system model is known, output-based design for nonlinear systems remains a difficult problem~\cite[Section~8.7]{bernard2022observer}. In particular, the certainty equivalence principle, which holds for linear systems, does not readily extend to the nonlinear case. 
	Moreover, when only output measurements are available, the number of state variables $n$ can be potentially unknown, thereby complicating the design of the reduction matrix. In addition, the interface function would typically have a dynamic output-feedback rather than a static state-feedback structure. Nevertheless, building on the results of~\cite{dai2023data,dai2025data} and leveraging the notion of uniform observability, one could explore this direction to extend the proposed framework. We note, however, that~\cite{dai2023data,dai2025data} consider mainly single-input single-output nonlinear systems.}
	
	\BS{According to~\eqref{eq:tmp13}, and in view of the term $\Vert \XdCm Q_1 \Vert$ in the cost function, the proposed framework aims to minimize the influence of the original system's nonlinearities and to capture their residual effect through $\eta$, which in turn motivates restricting ct-ROMs to linear dynamics. An alternative approach, potentially yielding tighter closeness guarantees, is to allow ct-ROMs to possess nonlinear dynamics and instead design the interface function so as to establish a relation between the nonlinear terms of the two systems (see~(A.4) in~\cite{lavaei2019compositional}), possibly resulting in $\mathbfcal{R}_2 \neq \bbzero_{(d - n) \times \hat n}$. The development of such extensions can also be pursued through efforts to bridge SF-based and moment-matching-based approaches~\cite[Section~III-D]{11312666}.}
	
	\BS{Two additional directions for future research include extending the framework to more general classes of nonlinear systems, potentially of the form $\dot{x} = f(x, u)$, as well as addressing noise affecting the state and/or input data in addition to the state-derivative measurements.}
	
	\section{Simulation Results}\phantomsection \label{Sec: simulation}
	In this section, we evaluate the effectiveness of our data-driven framework.
	\BS{To this end, we employ the proposed data-driven scheme to construct ct-ROMs for two nonlinear systems of dimensions $20$ and $12$, respectively, with unknown dynamics, encompassing complex nonlinearities, together with their corresponding interface functions. To enforce logical specifications over the (relatively) high-dimensional nonlinear systems, controllers are synthesized for the ct-ROMs using the formal-method tool \texttt{SCOTS}~\cite{rungger2016scots} that ensures satisfaction of the desired properties at the ct-ROM level. These results are then transferred to the original systems via the data-driven interface functions. Finally, we demonstrate that the observed output errors between the original systems and their ct-ROMs are consistent with the theoretical bounds, while the targeted logical specifications are successfully satisfied by the original systems. It is worth emphasizing that, even when the mathematical model of the original system is available, formal-method tools generally cannot accommodate systems with dimensions exceeding $5$ for formal controller synthesis due to computational complexity. This further highlights the importance of the proposed approach, enabling formal controller synthesis for (relatively) high-dimensional systems, even in the absence of explicit system models.}
	We performed all the simulations utilizing \textsc{Matlab} \textit{R2023b} on a MacBook Pro (Apple M2 Max, 32GB memory).
	
	\subsection{\BS{Inverter Chain: Safety Specification}}
	\BS{As the first case study, we consider a widely adopted benchmark in the MOR literature, whose model is borrowed from~\cite{bhattacharjee2025signal,morwiki_modNonRCL}. This benchmark consists of a circuit of chained inverter gates, commonly referred to as an inverter chain. The system evolves according to
		\begin{align*}
			\dot x = -x + \! \begin{bmatrix} 0\\ \mathrm{g}(x_1) \\ \vdots \\  \mathrm{g}(x_{n-1})\end{bmatrix} + \begin{bmatrix} u\\ 0 \\ \vdots \\  0 \end{bmatrix}\!\!,
		\end{align*}
		where $\mathrm{g}(x_i) = \nu \tanh(\mathrm{a}x_{i})$ for all $i \in \{1, \ldots, n-1\}$, with $\nu = 0.25$ denoting the supply voltage and $\mathrm{a} = 35$ representing a physical parameter. We consider $n = 20$, \emph{i.e.}, the system has $20$ state variables. In general, this system has a single control input, \emph{i.e.}, $m = 1$, which makes it particularly challenging to control when complex specifications must be satisfied. We note that the system dynamics are assumed to be unknown to us and are presented here solely for completeness. However, in accordance with Assumption~\ref{assump-on-D}, we assume that the dictionary $\mathcal{D}(x) = [x^\top ~ \mathrm{g}(x_1) ~ \ldots ~ \mathrm{g}(x_{19})]^\top$ is known.}
		
	\BS{Our goal is threefold: \emph{(i)} construct a ct-ROM with only one state variable (\emph{i.e.}, $\hat n = 1$) for the unknown original nonlinear system with $20$ state variables, \emph{(ii)} design a controller for the ct-ROM such that it satisfies a given safety specification, and \emph{(iii)} refine this controller back to the original system through a data-driven interface design, thereby ensuring that the original system satisfies the safety specification.
	To this end, we follow Algorithm~\ref{Alg:1} and first collect input--state data from the system, as in~\eqref{eq: data_ct}, with $T = 700$ as the number of collected samples. During data collection, the initial condition of each state variable is randomly selected from the interval $[-1,1]$, and the input signal is chosen as a random sequence taking values in $\{1,-1\}$ at each sampling instant. We assume that the noise associated with the state-derivative data at each sampling time lies within the interval $[-0.001,\,0.001]$. This implies that Assumption~\ref{assump:noise} is satisfied with $\mathcal{E}\mathcal{E}^\top = 0.0140 \I_{20}$. We also note that the collected data satisfy Assumption~\ref{assumpt:full-rank}.}

    \begin{figure}[t!]
		\centering
		\begin{subfigure}[b]{\linewidth}
			\centering
			\includegraphics[width=0.75\linewidth]{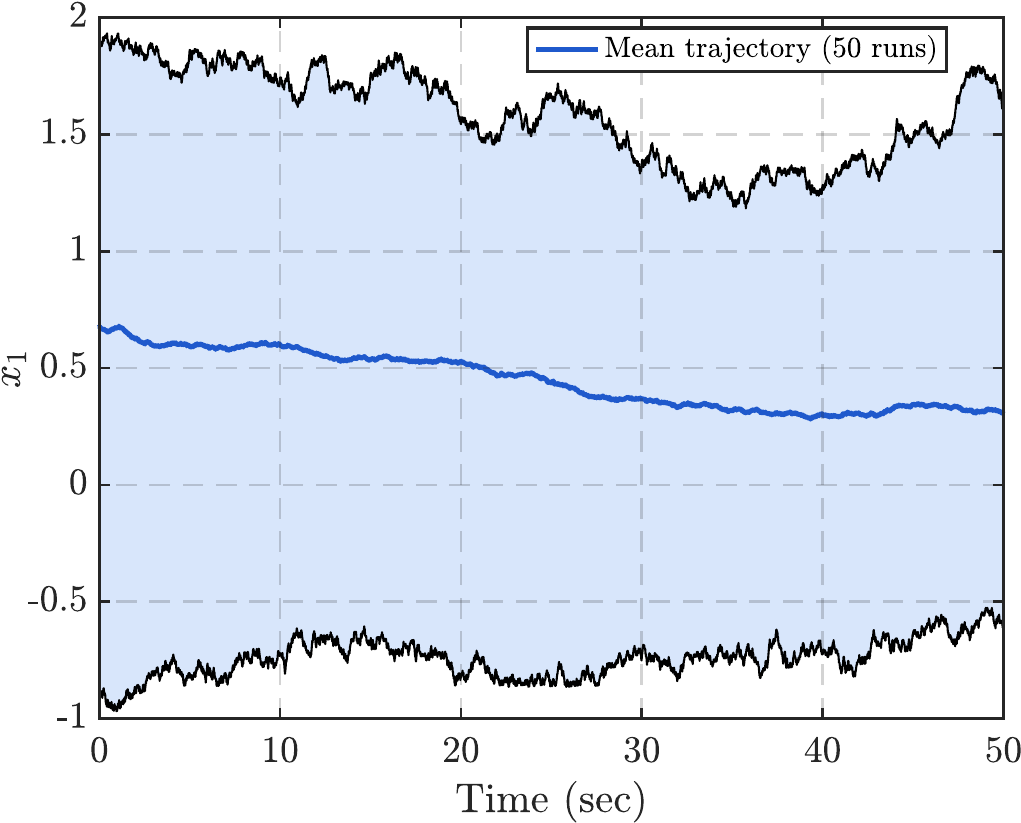}
			\caption{Safety envelope and the mean trajectory}
			\label{fig:subfig1_1}
		\end{subfigure}\vspace{0.4cm}

		\begin{subfigure}[b]{\linewidth}
			\centering
			\includegraphics[width=0.75\linewidth]{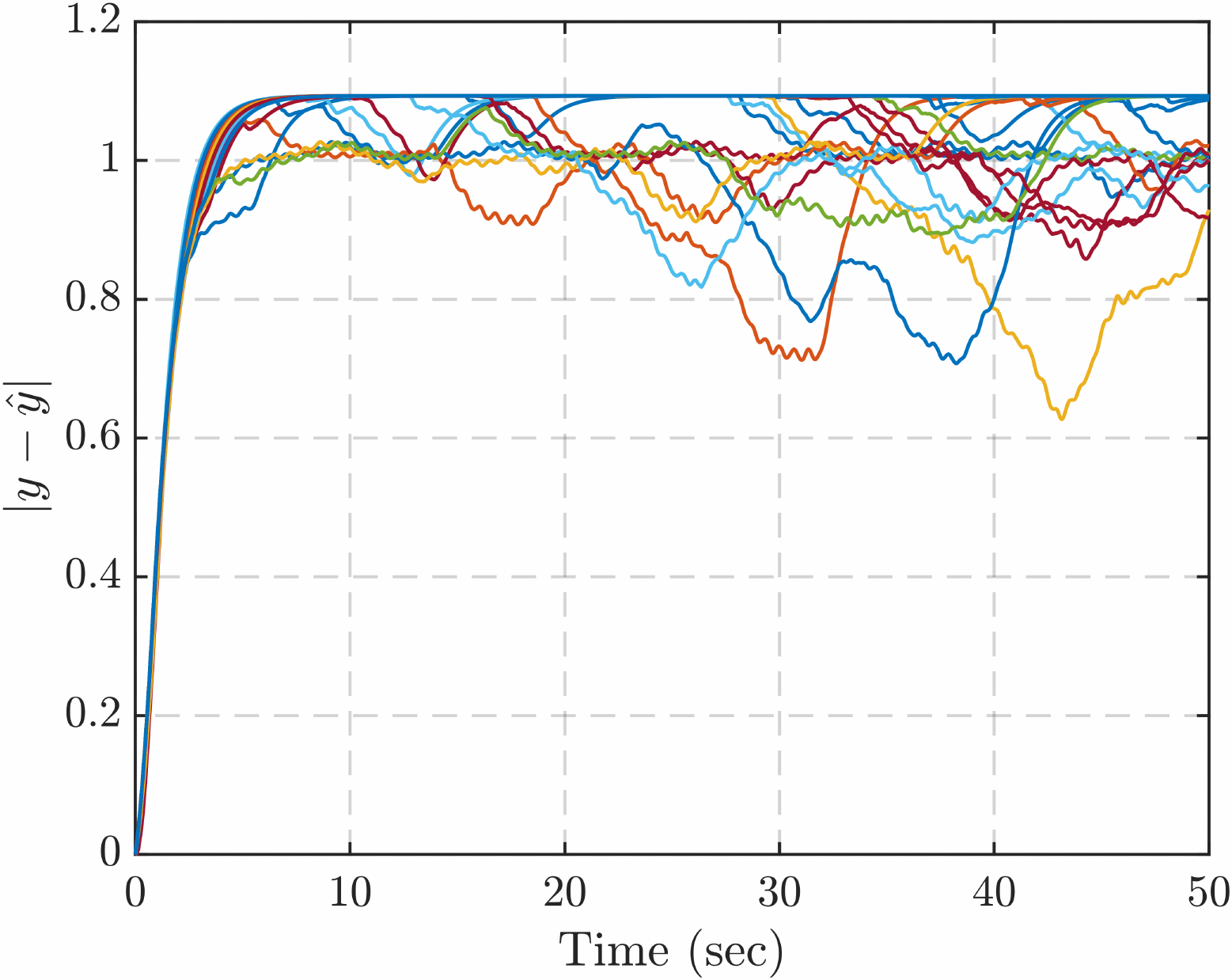}
			\caption{Output trajectory closeness between two systems per experiment}
			\label{fig:subfig1_2}
		\end{subfigure}
		
		\caption{\BS{(a) The mean trajectory and the safety envelope over $50$ simulation runs with arbitrary initial conditions. The blue curve represents the mean state trajectory, while the black curves denote the pointwise minimum and maximum across all trajectories at each time instant. The results show that the system trajectories remain within the prescribed safety bounds for all runs. (b) The error between the output trajectories of the two systems in each simulation run, showing that the obtained closeness guarantee is not violated.}
		}
		\label{fig: example 1}
	\end{figure}
	
	\BS{According to Step~\ref{Step3} in Algorithm~\ref{Alg:1}, we select $\hat{A} = -0.01$ and $\hat B = 1$, and set $\kappa = 0.7$. With these choices, the SDP~\eqref{eq: SDP} is feasible and returns the parameters required for constructing the interface function and computing the closeness guarantee~\eqref{eq: error-cont}. Since the resulting matrices are very large (as $T = 700$ and $n = 20$), we do not report them explicitly; instead, we provide other relevant quantities: $\mu_1 = 1$, $\mu_j = 0.1$ for all $j \in \{2, \ldots, 5\}$, $\mu_6 = 85.4275$, $\Vert Q_1 \Vert = 0.3210$, $\Vert \XdCm Q_1 \Vert = 0.25$, $\Vert Q_2 \Vert = 0.3301$, $\Vert Q_3 \Vert = 0.0390$, $\Vert \XdCm Q_3  -  \XC Q_2 \hat B \Vert = 7.2716 \times 10^{-4}$, $\Xi = 0.99$, $\Psi = 0.9999$, and $\alpha = 0.3457$. With these values, we compute $\rho = 2.1807 \times 10^{-4}$. Moreover, one can obtain $\max_{x \in X} \vert \mathcal{N}(x) \vert^2 = n - 1 = 19$, independently of the choice of $X$. For the considered safety specification, we require that $x_1$ always remains within $[-1,2]$. Accordingly, we impose the same safety specification for the ct-ROM and compute $\max_{\hat x \in \hat X} \vert \hat x \vert^2 = 4$. Moreover, we consider $\hat u \in [-1, 1]$. Collectively, these lead to $\eta = 1.5226$. Consequently, the closeness guarantee~\eqref{eq: error-cont} is computed as $2.5084$, with $\CTSF(x_0, \hat x_0) = 0$ (cf., Remark~\ref{rem: eta_and_error}). The corresponding simulation results are presented in Fig.~\ref{fig: example 1}. As shown in Fig.~\ref{fig:subfig1_1}, the safety specification is satisfied across $50$ experiments, each initialized from a distinct initial condition. Moreover, Fig.~\ref{fig:subfig1_2} demonstrates that the closeness guarantee is never violated.}
	
	\subsection{\BS{Academic System: Recurrence Specification}}
	\BS{While the preceding case study demonstrates the applicability of the proposed framework to practical systems with relatively high dimensions, the second case study aims to show that rich LTL specifications can also be handled smoothly within the framework. Moreover, the considered academic example exhibits highly nonlinear dynamics, rendering the problem more challenging. In particular, the system evolves according to the following dynamics:
	\begin{align*}
		&\dot{x}_1 \! = \! - 2x_1 \! - \! x_2 - \! 0.25 \ln(1  \! +\! x_{10}^2) \! + \!  \frac{0.7\ln(1 \! + \! \tanh^4 (x_{1})\!)}{1 \! + \! x^2_{5}} + u_1,\\
		& \dot{x}_2 = 2x_1 - 3x_2 - 2x_3,\\
		&\dot{x}_3 = x_2 - 4x_3 - 2x_4 + 0.2 \tan^{-1}(\sin^2(x_8 x_{10})),\\
		&\dot{x}_4 = 2x_5 - 5x_4 - x_3 + u_2,\\
		&\dot{x}_5 = - 6x_5 - x_6,\\
		&\dot{x}_6 = 2x_7 - 7x_6 - 2x_5,\\
		&\dot{x}_7 = x_6 - 8x_7 - 2x_8,\\
		&\dot{x}_8 = - x_7 - 9x_8 - \frac{0.5 \sin^2(x_1)}{1 + x_3^2} + u_3,\\
		&\dot{x}_9 = - 2x_8 - 10x_9 - 2x_{10},\\
		&\dot{x}_{10} = x_9 - 11x_{10} + 0.25 \ln(1 + x^2_{10}) + u_4,\\
		&\dot{x}_{11} = x_{10} - 12x_{11},\\
		&\dot{x}_{12} = x_{11} - 13x_{12}.
		\end{align*}
     We assume that the above dynamics describe a shuttle taxi that must repeatedly navigate between two places infinitely often (\emph{i.e.,} a recurrence property in temporal logic, which is inherently complex) while avoiding obstacles along the way. Notably, the corresponding dynamics contain an unmatched nonlinearity in $x_3$ (\emph{i.e.,} no control input directly influences this state and therefore its effect cannot be directly canceled by the control input), which further complicates the problem and highlights the effectiveness of the proposed framework.
	We emphasize that the dynamics are assumed to be unknown to us and are presented here solely for completeness. However, we assume that a complete dictionary satisfying Assumption~\ref{assump-on-D} is available.}
	
	\BS{Our goal is similar to the preceding case study, with two main differences: \emph{(i)} we aim to construct a ct-ROM with two state variables (\emph{i.e.}, $\hat{n} = 2$) for the original unknown system with $12$ state variables, and \emph{(ii)} the specification of interest is a recurrence task. Accordingly, we follow Algorithm~\ref{Alg:1} to achieve this goal. Specifically, we collect input--state data from the system with $T = 50$, satisfying Assumption~\ref{assumpt:full-rank}. For the data-gathering experiment, the initial condition is chosen as $\One_{12}$. Moreover, we assume that the noise associated with the state-derivative data at each sampling time lies within the interval $[-0.002,\,0.002]$. Consequently, Assumption~\ref{assump:noise} holds with $\mathcal{E}\mathcal{E}^\top = 0.0024\,\I_{12}$.}
	
\begin{figure}[t!]
		\centering
		\begin{subfigure}[b]{\linewidth}
			\centering
			\includegraphics[width=0.7\linewidth]{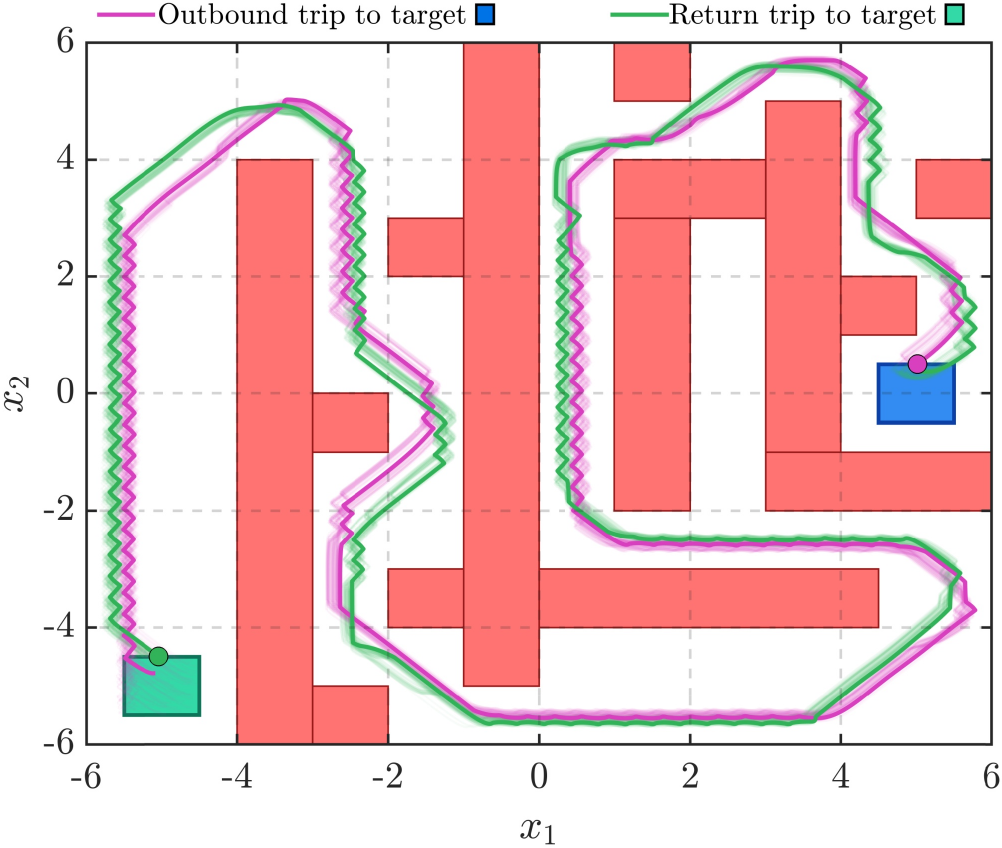}
			\caption{$100$ trajectories of original system with arbitrary initial conditions}
			\label{fig:subfig2_1}
		\end{subfigure}\vspace{0.4cm}

		\begin{subfigure}[b]{\linewidth}
			\centering
			\includegraphics[width=0.7\linewidth]{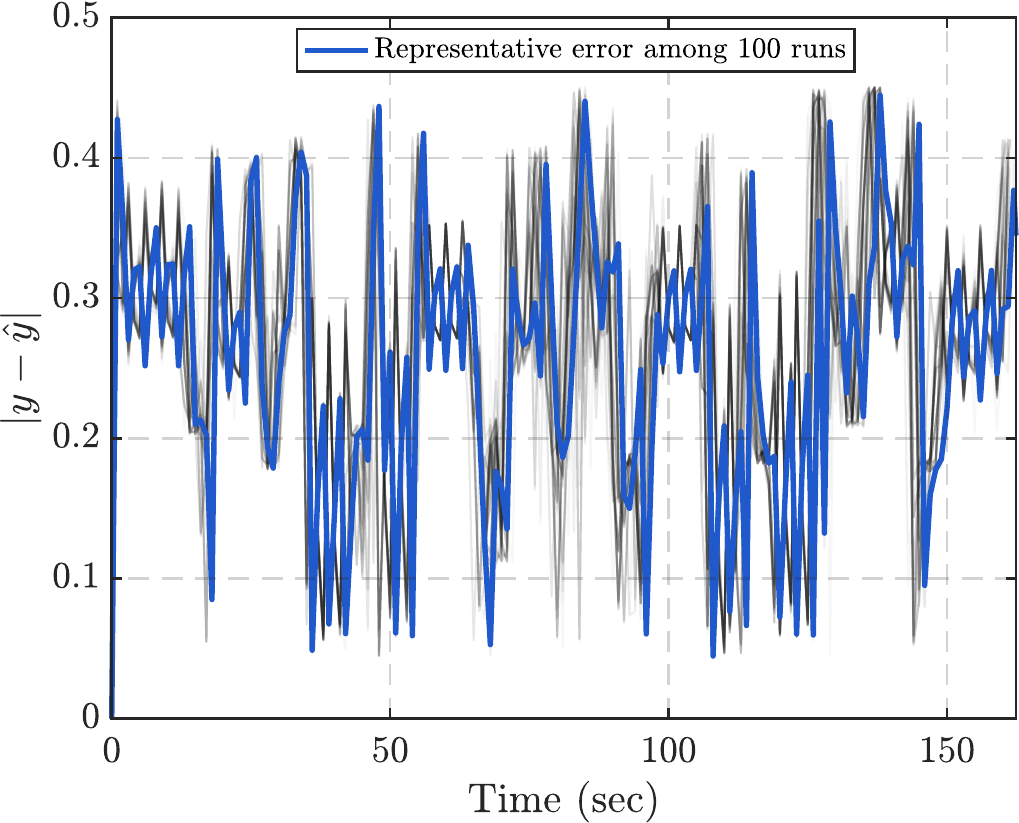}
			\caption{Output trajectory closeness between two systems per experiment}
			\label{fig:subfig2_2}
		\end{subfigure}\vspace{0.4cm}
		
		\begin{subfigure}[b]{\linewidth}
			\centering
			\includegraphics[width=0.7\linewidth]{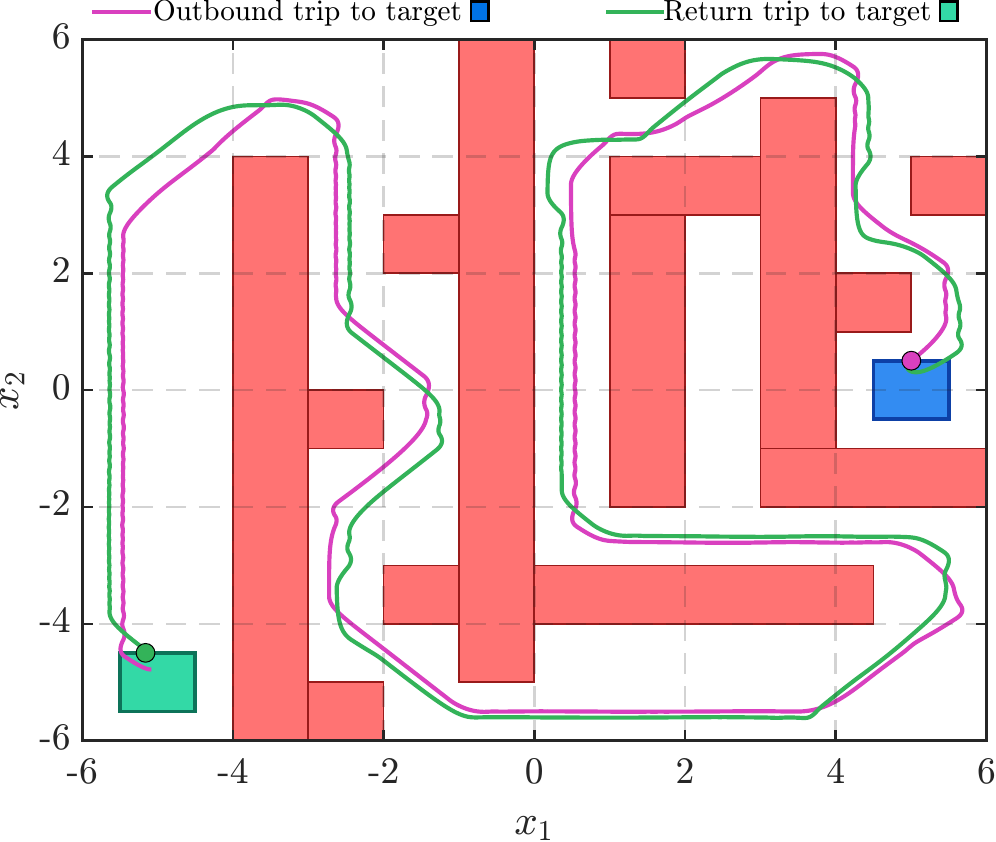}
			\caption{System trajectory after applying a smoothing filter}
			\label{fig:subfig2_3}
		\end{subfigure}
		
		\caption{\BS{(a) The trajectories of the original system for $100$ outbound trips starting from different initial conditions \legendsquare{START} to the target \legendsquare{TARGET} and $100$ return trips from \legendsquare{TARGET} to \legendsquare{START}, while avoiding the obstacles \legendsquare{OBSTACLES}. One representative outbound trip and one representative return trip are depicted in bold, whereas the rest are shown in a faded style. (b) The error between the output trajectories of the two systems in each simulation run, illustrating that the obtained closeness guarantee is not violated.
				(c) A representative system trajectory after applying a smoothing filter to attenuate the zigzag behavior observed in (a), stemming from the \texttt{SCOTS} synthesis tool, due primarily to the chosen discretization parameter. This simulation's video is available at \url{https://youtu.be/3N1B1E1GQIs}.}
		}
		\label{fig: example 2}
	\end{figure}

	\BS{Following Step~\ref{Step3} of Algorithm~\ref{Alg:1}, we choose
	\(\hat{A} = -0.0001 \I_2\) and \(\hat B = 0.1 \I_2\),
	and fix $\kappa = 2.3$. Subsequently, the SDP~\eqref{eq: SDP} is feasible and yields the coefficients required to construct the interface function as well as to quantify the closeness guarantee~\eqref{eq: error-cont}. Accordingly, we achieve $\mu_j = 2$ for all $j \in \{1,\ldots,5\}$, $\mu_6 = 171.7933$, $\Vert Q_1 \Vert = 2.7623$, $\Vert \XdCm Q_1 \Vert = 0.1815$, $\Vert Q_2 \Vert = 1.9837$, $\Vert Q_3 \Vert = 0.0119$, $\Vert \XdCm Q_3 - \XC Q_2 \hat{B} \Vert = 0.2010$, and $\alpha = 0.3602$. These values lead to the bound $\rho = 0.0202$. Furthermore, considering $x_{10} \in [-6, 6]$ and regardless of the other state variables, it can be shown that $\max_{x \in X} \vert \mathcal{N}(x) \vert^2 \leq 16$ (cf., Remark~\ref{rem:eta_comp}). We also consider that $\hat X = [-6, 6]^2$, yielding $\max_{\hat{x} \in \hat{X}} \vert \hat{x}\vert^2 = 72$, while restricting the input to $\hat{u} \in [-6,6]^2$. These selections collectively result in $\eta = 0.7502$. Consequently, the closeness bound in~\eqref{eq: error-cont} is computed as $1.6314$ with $\CTSF(x_0,\hat{x}_0)=0$.
	The corresponding simulation results are depicted in Fig.~\ref{fig: example 2}. As illustrated in Fig.~\ref{fig:subfig2_1}, the complex recurrence specification is satisfied across $100$ simulation runs, while Fig.~\ref{fig:subfig2_2} confirms that the derived closeness guarantee is never violated.
	}
	
	\section{Conclusion}\phantomsection \label{Sec: conclusion}
	In this paper, we introduced a data-driven methodology for constructing ROMs of nonlinear dynamical systems, when explicit mathematical representations were unavailable. By utilizing the concept of SFs, a formal relation was established between the output trajectories of the original systems and those of their data-driven ROMs, ensuring a well-defined measure of closeness. 
	\BS{To facilitate this process, we collected one set of noise-corrupted input--state data from the system and proposed conditions taking the form of data-dependent SDPs to construct both ROMs and SFs.} We demonstrated that the constructed ROMs can be leveraged for control synthesis, enabling the enforcement of high-level logic properties over the unknown system. This was achieved by first designing controllers for the ROMs using data and subsequently mapping the control strategies back to the original system via a data-driven interface function. The effectiveness of our proposed methodology was validated through {two} case studies with nonlinear dynamics, {one of which was a practical circuit with $20$ state variables}.
	Future research will focus on extending the framework to construct nonlinear ROMs, expanding the class of systems under consideration, and utilizing input--output data.
	
	\bibliographystyle{IEEEtran}
	\bibliography{biblio}

	\begin{IEEEbiography}[{\includegraphics[width=1in,height=1.3in,clip,keepaspectratio]{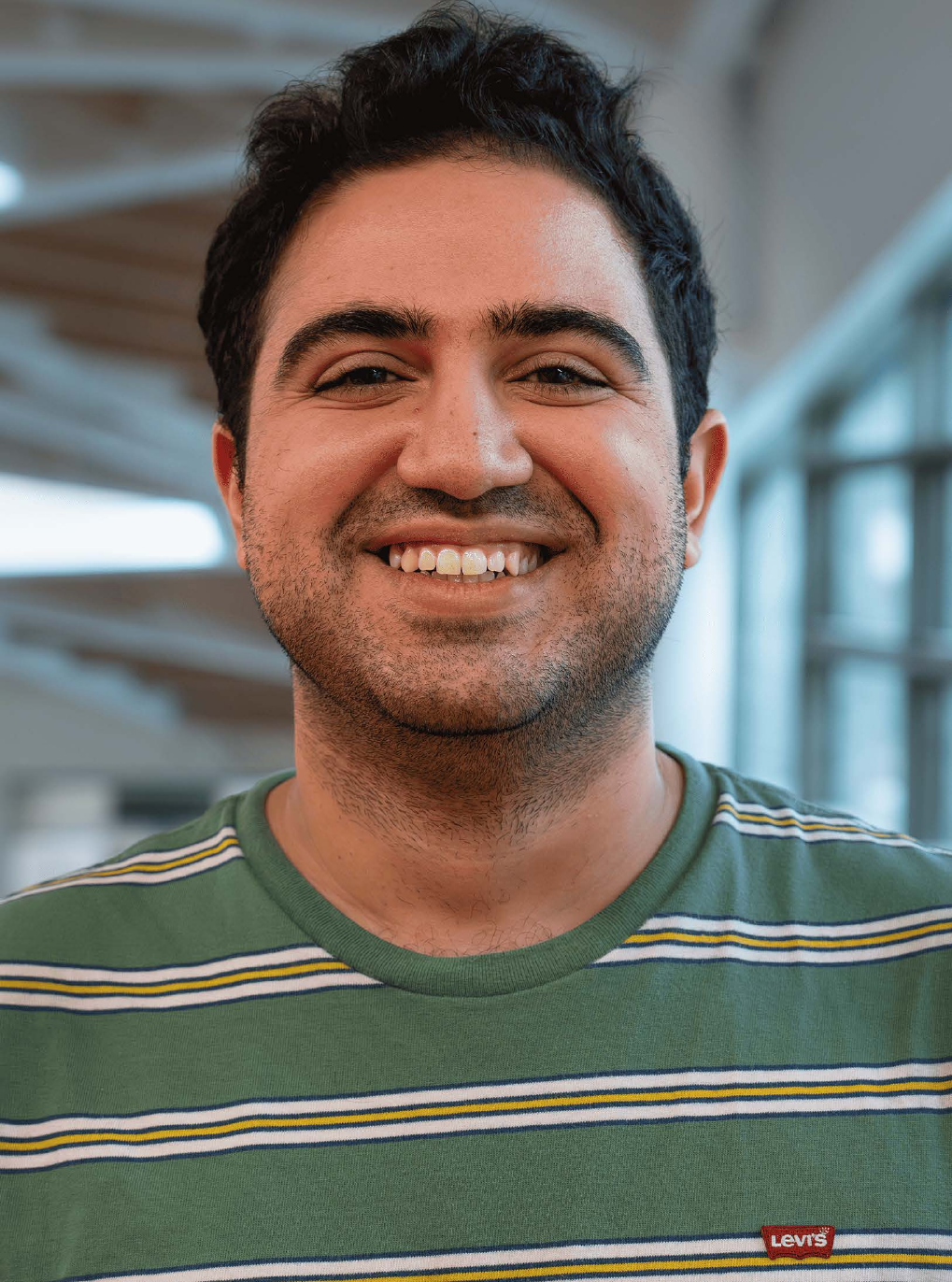}}]{Behrad Samari}~(Student Member, IEEE) received his B.Sc. and M.Sc. degrees in electrical engineering, control major, from K. N. Toosi University of Technology, Tehran, Iran, and University of Tehran (UT), Tehran, Iran, in 2019 and 2022, respectively. He is currently pursuing his PhD in the School of Computing at Newcastle University, U.K. He is the Best Repeatability Prize Finalist at the 8$^{\text{th}}$ IFAC Conference on Analysis and Design of Hybrid Systems (ADHS), 2024. His research interests include (nonlinear) control and system theory, data-driven approaches, and formal methods.
	\end{IEEEbiography}

	\begin{IEEEbiography}[{\includegraphics[width=1in,height=1.3in,clip,keepaspectratio]{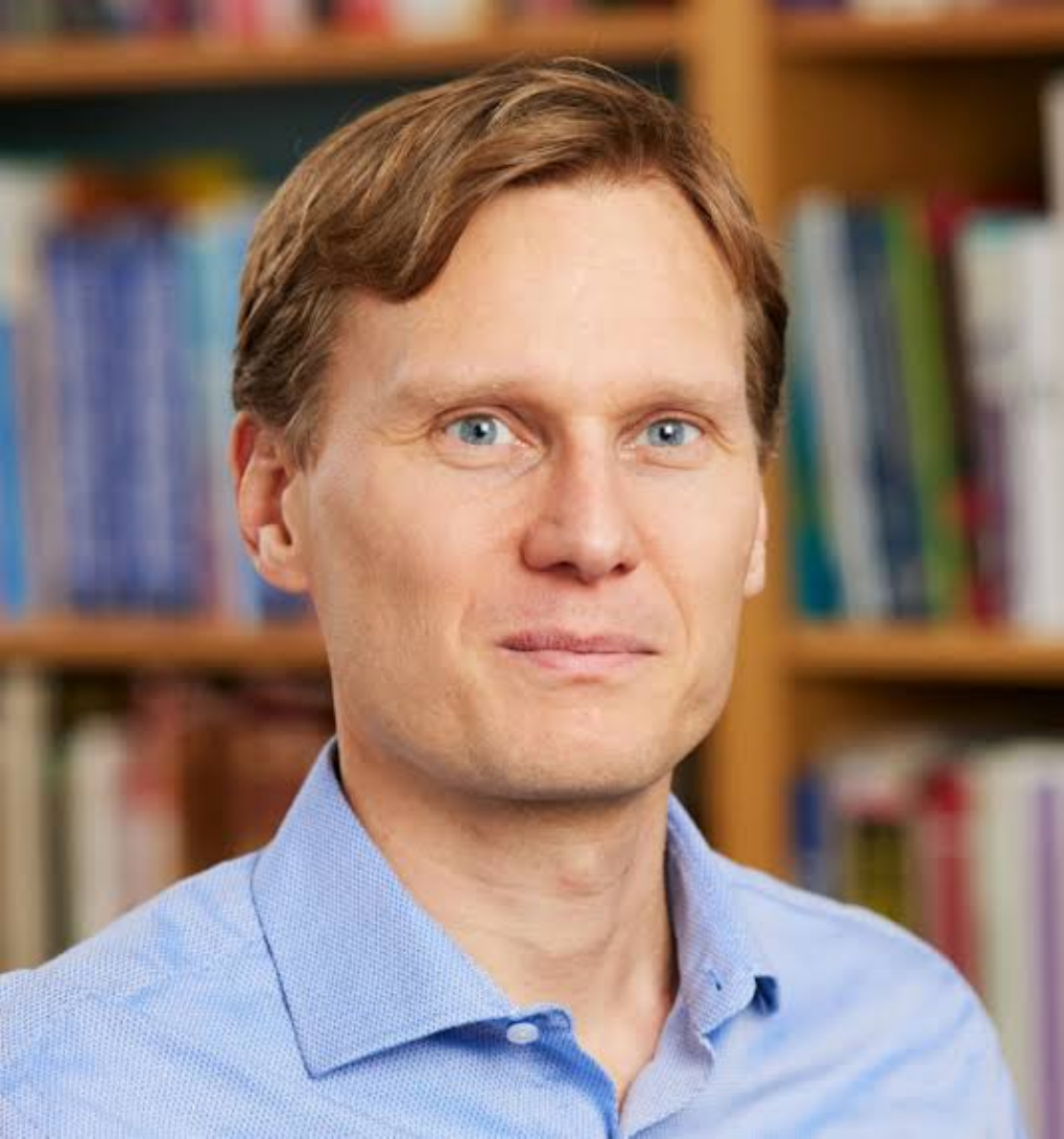}}]{Henrik Sandberg}
		(Fellow, IEEE) is Professor at the Division of Decision and Control Systems, KTH Royal Institute of Technology, Stockholm, Sweden. He received the M.Sc. degree in engineering physics and the Ph.D. degree in automatic control from Lund University, Lund, Sweden, in 1999 and 2004, respectively. From 2005 to 2007, he was a Postdoctoral Scholar at the California Institute of Technology, Pasadena, USA. In 2013, he was a Visiting Scholar at the Laboratory for Information and Decision Systems (LIDS) at MIT, Cambridge, USA. He has also held visiting appointments at the Australian National University and the University of Melbourne, Australia. His current research interests include security of cyber-physical systems, power systems, model reduction, and fundamental limitations in control. Dr. Sandberg was a recipient of the Best Student Paper Award from the IEEE Conference on Decision and Control in 2004, an Ingvar Carlsson Award from the Swedish Foundation for Strategic Research in 2007, and a Consolidator Grant from the Swedish Research Council in 2016. He has served on the editorial boards of IEEE Transactions on Automatic Control and the IFAC Journal Automatica, and is currently an elected member of the IEEE Control Systems Society Board of Governors. He is Fellow of the IEEE.
	\end{IEEEbiography}

	\begin{IEEEbiography}[{\includegraphics[width=1in,height=1.3in,clip,keepaspectratio]{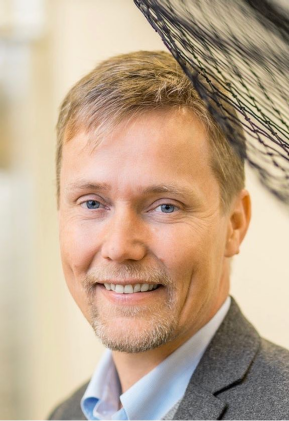}}]{Karl H. Johansson}~(Fellow, IEEE) is Swedish Research Council Distinguished Professor in Electrical Engineering and Computer Science at KTH Royal Institute of Technology in Sweden and Founding Director of Digital Futures. He earned his MSc degree in Electrical Engineering and PhD in Automatic Control from Lund University. He has held visiting positions at UC Berkeley, Caltech, NTU and other prestigious institutions. His research interests focus on networked control systems and cyber-physical systems with applications in transportation, energy, and automation networks. For his scientific contributions, he has received numerous best paper awards and various other distinctions from IEEE, IFAC, and other organizations. He has been awarded Distinguished Professor by the Swedish Research Council, Wallenberg Scholar by the Knut and Alice Wallenberg Foundation, Future Research Leader by the Swedish Foundation for Strategic Research. He has also received the triennial IFAC Young Author Prize, IEEE CSS Distinguished Lecturer, IFAC Outstanding Service Award, and IEEE CSS Hendrik W. Bode Lecture Prize. His extensive service to the academic community includes being President of the European Control Association, IEEE CSS Vice President Diversity, Outreach \& Development, and Member of IEEE CSS Board of Governors and IFAC Council. He has served on the editorial boards of Automatica, IEEE TAC, IEEE TCNS and many other journals. He has also been a member of the Swedish Scientific Council for Natural Sciences and Engineering Sciences. He is Fellow of both the IEEE and the Royal Swedish Academy of Engineering Sciences.
	\end{IEEEbiography}

	\begin{IEEEbiography}[{\includegraphics[width=1in,height=1.25in,clip,keepaspectratio]{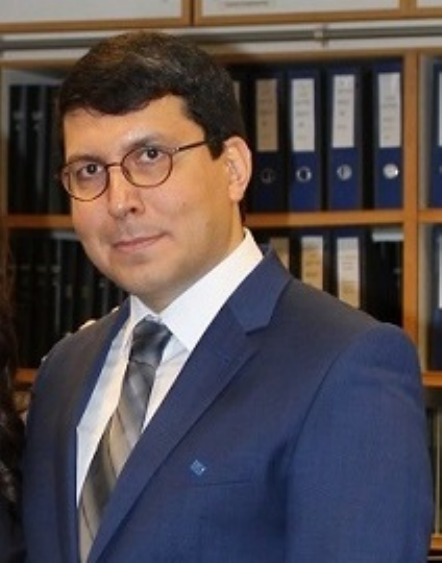}}]{Abolfazl Lavaei}~(M'17--SM'22) is an Assistant Professor in the School of Computing at Newcastle University, United Kingdom. Between January 2021 and July 2022, he was a Postdoctoral Associate in the Institute for Dynamic Systems and Control at ETH Zurich, Switzerland. He was also a Postdoctoral Researcher in the Department of Computer Science at LMU Munich, Germany, between November 2019 and January 2021. He received the Ph.D. degree in Electrical Engineering from the Technical University of Munich (TUM), Germany, in 2019. He obtained the M.Sc. degree in Aerospace Engineering with specialization in Flight Dynamics and Control from the University of Tehran (UT), Iran, in 2014. He is the recipient of several international awards in the acknowledgment of his work including ADHS Best Repeatability Prize (Finalist) 2024 and 2021, HSCC Best Demo/Poster Awards 2022 and 2020, IFAC Young Author Award Finalist 2019, and Best Graduate Student Award 2014 at University of Tehran with the full GPA (20/20). His research interests revolve around the intersection of Control Theory, Formal Methods, and Statistical Learning Theory.
	\end{IEEEbiography}
	
\end{document}